\documentclass[universe,article,accept,moreauthors]{Definitions/mdpi}
\usepackage{amsmath}
\usepackage{amsfonts}
\usepackage{amssymb,bm}
\usepackage{siunitx}
\usepackage{color}
\usepackage{braket}

\usepackage{upgreek}
 \usepackage{textcomp}

\firstpage{1}
\pubvolume{7}
\issuenum{4}
\articlenumber{93}
\pubyear{2021}
\copyrightyear{2020}
\history{Received: 21 March 2021; Accepted: 4 April 2021; Published: 7 April 2021}

\Title{ Measurement of the Casimir Force between 0.2 and 8~$\upmu$m:
Experimental Procedures and Comparison with Theory  }

\Author{Giuseppe Bimonte $^{1,2}$,
Benjamin Spreng $^{3,4}$,
Paulo A. Maia Neto $^{5}$, Gert-Ludwig Ingold $^{3}$,
Galina L. Klimchitskaya $^{6,7}$, Vladimir M. Mostepanenko $^{6,7,8}$,
Ricardo S. Decca $^{9}$}

\address{%
$^{1}$ \quad  Dipartimento di Fisica E. Pancini, Universit\`{a} di
Napoli Federico II, Complesso Universitario
di Monte S. Angelo,  Via Cintia, I-80126 Napoli, Italy\\
$^{2}$ \quad INFN Sezione di Napoli, I-80126 Napoli, Italy\\
$^{3}$ \quad Universit{\"a}t Augsburg, Institut f{\"u}r Physik, 86135 Augsburg, Germany\\
$^{4}$ \quad Department of Electrical and Computer Engineering, University of California, Davis, California 95616, USA\\
$^{5}$ \quad Instituto de F{\'i}sica, Universidade Federal do Rio de Janeiro, CP 68528, Rio de Janeiro RJ 21941-909, Brazil\\
$^{6}$ \quad Central Astronomical Observatory at Pulkovo of the Russian Academy of Sciences, Saint Petersburg, 196140, Russia\\
$^{7}$ \quad Institute of Physics, Nanotechnology and
Telecommunications, Peter the Great Saint Petersburg
Polytechnic University, Saint Petersburg, 195251,  Russia\\
$^{8}$ \quad Kazan Federal University, Kazan, 420008, Russia\\
$^{9}$ \quad Department of Physics, Indiana University-Purdue University Indianapolis, Indianapolis, IN 46202, USA}

\corres{Correspondence: rdecca@iupui.edu}

\abstract{We present results on the determination of the
differential Casimir force between an Au-coated sapphire sphere
and the top and bottom of Au-coated deep silicon trenches
performed by means of the micromechanical torsional oscillator
in the range of separations from 0.2 to 8 $\mu$m. The random and
systematic errors in the measured force signal are determined
at the 95\% confidence level and combined into the total
experimental error. The role of surface roughness and edge effects
is investigated and shown to be negligibly small. The distribution
of patch potentials is characterized by Kelvin probe microscopy,
yielding an estimate of the typical size of patches,
the respective r.m.s. voltage and their impact on the measured
force. A comparison between the experimental results and theory
is performed with no fitting parameters. For this purpose, the
Casimir force in the sphere-plate geometry is computed independently
on the basis of first principles of quantum electrodynamics
using the scattering theory and the gradient expansion. In doing
so, the frequency-dependent dielectric permittivity of Au is
found from the optical data extrapolated to zero frequency by
means of the plasma and Drude models. It is shown that the
measurement results exclude the Drude model extrapolation over the
region of separations from 0.2 to 4.8~$\mu$m, whereas the alternative
extrapolation by means of the plasma model is experimentally
consistent over the entire measurement range. A discussion of the
obtained results is provided.}


\begin{document}
\section{Introduction}\label{Intro}

The Casimir attraction \cite{1} between two uncharged closely
spaced bodies is a macroscopic quantum effect which is caused
by the zero-point and thermal fluctuations of the electromagnetic
field. Over a long period of time, it was measured only up to
an order of magnitude. The modern period in experimental
investigation of this phenomenon started from measuring
the Casimir force between the Au-coated surfaces of a large
lens and a plate by means of the torsion pendulum \cite{1a}.
Precise measurements of the Casimir force have been made
possible only during the last 20 years thanks to microtechnology
achievements. These measurements gave the possibility to
quantitatively check the theoretical predictions of the Lifshitz
theory \cite{2,3} which generalizes the original Casimir prediction
(made for two parallel ideal metal planes at zero temperature) for
the case of thick plates described by their frequency-dependent
dielectric permittivities in thermal equilibrium with the
environment.

The first experiment of this kind used an atomic force microscope
where the sharp tip was replaced with the sphere of about
100 $\mu$m radius \cite{4}. This experiment made it possible to
demonstrate corrections to the famous Casimir expression due to
the finite conductivity of the boundary metal. Another novel
facility of nanotechnology, a micromechanical torsional oscillator,
was used to demonstrate the actuation of micromechanical devices
by the Casimir force \cite{5,6}. After experimental improvements
\cite{7}, micromechanical torsional oscillators were used in the
most precise measurements of the Casimir interaction between an
Au-coated sapphire sphere of 150 $\mu$m radius and an Au-coated
polysilicon plate \cite{8,9,10,11}.

It turned out that the theoretical predictions
of the Lifshitz theory are excluded by
the measurement data  \cite{8,9,10,11} if the dielectric permittivity is
obtained from the measured optical data of Au \cite{12}
extrapolated down to zero frequency by means of the
dissipative Drude model which takes the proper account of the
relaxation properties of conduction electrons.
The Lifshitz theory using the Drude model was excluded over
the separation region from 160 to 750 nm. What is even more
surprising, an agreement between experiment and theory was
restored  \cite{8,9,10,11} if, except of the Drude model, an extrapolation was
made using the dissipationless plasma model which disregards
the relaxation properties of conduction electrons and should
be applicable only at sufficiently high frequencies of
infrared optics.
Similar results were obtained later in experiments
using an atomic force microscope for both Au surfaces \cite{13}
and for the surfaces of a sphere and a plate coated with layers
of magnetic metal Ni \cite{14,15}.

In 2016, following the proposal of \cite{16}, the isoelectronic
experiment was performed \cite{17} on measuring the differential
Casimir force between a Ni-coated sphere and Ni and Au sectors
of a rotating disc coated with an Au overlayer. This
configuration vastly enhances the role of the zero-frequency
term in the Lifshitz formula whose value mostly depends on
whether the Drude or the plasma model is used for an extrapolation
of the optical data. In so doing the theoretical predictions using
these models differ by up to a factor of 1000. As a result, the
Lifshitz theory using the Drude model was conclusively excluded
by the measurement data over the separation region from 200 to
700 nm. The same theory using the plasma model was found to be
in good agreement with the data over the entire measurement range.
In later experiments using an atomic force microscope and the
UV- and Ar-ion cleaned Au surfaces of a sphere and a plate, an
exclusion of the Drude model was confirmed up to the
separation distance of 1.1 $\mu$m \cite{18,19,20}.

In view of the fact that at separations $z\lesssim 1~\mu$m the
experimental results for metallic surfaces are already completely
settled, special attention is now attracted to the separation
region from 1 $\mu$m to a few micrometers. An upgraded
version of the experiment \cite{1a}, which measures the force between
an Au-coated sphere of centimeter-size curvature radius spaced
above an Au-coated plate in this separation range, was performed using
a torsion pendulum \cite{21}. The immediately measured quantity was not the
Casimir force, but up to an order of magnitude larger force
presumably determined by the surface patches on the Au surfaces.
The distribution of patch potential, and hence the corresponding
electrostatic force contribution to the measured force, was
not determined. The Casimir force was extracted from
the data by fitting with
two fitting parameters (the r.m.s. voltage fluctuations
over the surfaces and the force offset due to the
voltage offset in the setup electronics). The obtained results
were found to be in better agreement with the Drude model \cite{21}.
In \cite{22} it was argued that this conclusion is unjustified due
to the role of unavoidable imperfections (bubbles, pits, and
scratches) which are present on the surfaces of macroscopic
lenses. Moreover, according to the results of \cite{23}, at
separations exceeding 3 $\mu$m the measurement data of \cite{21}
subjected to the same fitting procedure are in better agreement
not with the Drude but with the plasma model.

In this article, we report measurements of the
differential Casimir force between an Au-coated sapphire sphere
and the top and bottom of Au-coated deep Si trenches in the
separation region from 0.2 to 8 $\mu$m. The measurements are
performed in vacuum by means of the micromechanical torsional
oscillator using a similar approach to those described in \cite{17,24}.
Taking into account the deepness of the trenches and the
thicknesses of Au coatings on both test bodies, the measured
quantity is the Casimir force acting between an all-gold sphere
and an all-gold plate.

The profiles of interacting surfaces were investigated by means
of an atomic force microscope with a sharp tip and the r.m.s.
roughness was determined. An impact of roughness on the Casimir
force turned out to be negligible. The random and systematic
errors in the measured Casimir force are found at the 95\%
confidence level and combined into the total experimental errors.
The edge effects, i.e., possible deviations of the form of
measured force signal from a Heaviside step function are
analyzed and found to be negligible.

Special attention is paid to the effect of patch potentials.
For this purpose, the interacting surface was characterized
by Kelvin probe microscopy and the typical sizes of patches and
respective r.m.s. voltage were determined. Using the theoretical
approach of \cite{25}, this allowed an estimation of the
attractive force originating from the surface patches which was
included in the balance of errors and uncertainties.

The experimental data were compared with no fitting parameters
with theoretical predictions for the Casimir force in
the sphere-plate geometry calculated numerically using the
scattering approach in the plane-wave basis \cite{26,27,28}
and the gradient
expansion \cite{29,30,31}. In doing so the dielectric
permittivity of Au was obtained from the tabulated optical data
extrapolated to zero frequency by means of the Drude or the
plasma model. It is shown that the theoretical predictions
using the Drude model for extrapolation of the optical data
are excluded by the measurement results within the range of
separations from 0.2 to 4.8 $\mu$m. The same theory using an
extrapolation by means of the plasma model is found to be in
agreement with the data over the entire measurement range.

The article is organized as follows: In Section 2, the details
of the experimental setup and the measurement procedures are
presented. In Section 3, we determine the sources of systematic
errors and evaluate the role of electrostatic patches and edge
effects. Sections 4 and 5 contain calculation of the Casimir force
between a sphere and a plate using the scattering approach and
the gradient expansion, respectively. The random errors are determined in
Section 6, combined with the systematic ones and used in the
comparison between experiment and theory. Section 7 contains
a discussion of the obtained results. In Section 8, the reader
will find our conclusions.

\section{Materials, Methods and Results}\label{MM}

\begin{figure}[!t]
	\centering
	\includegraphics[width=10.5cm,angle=270]{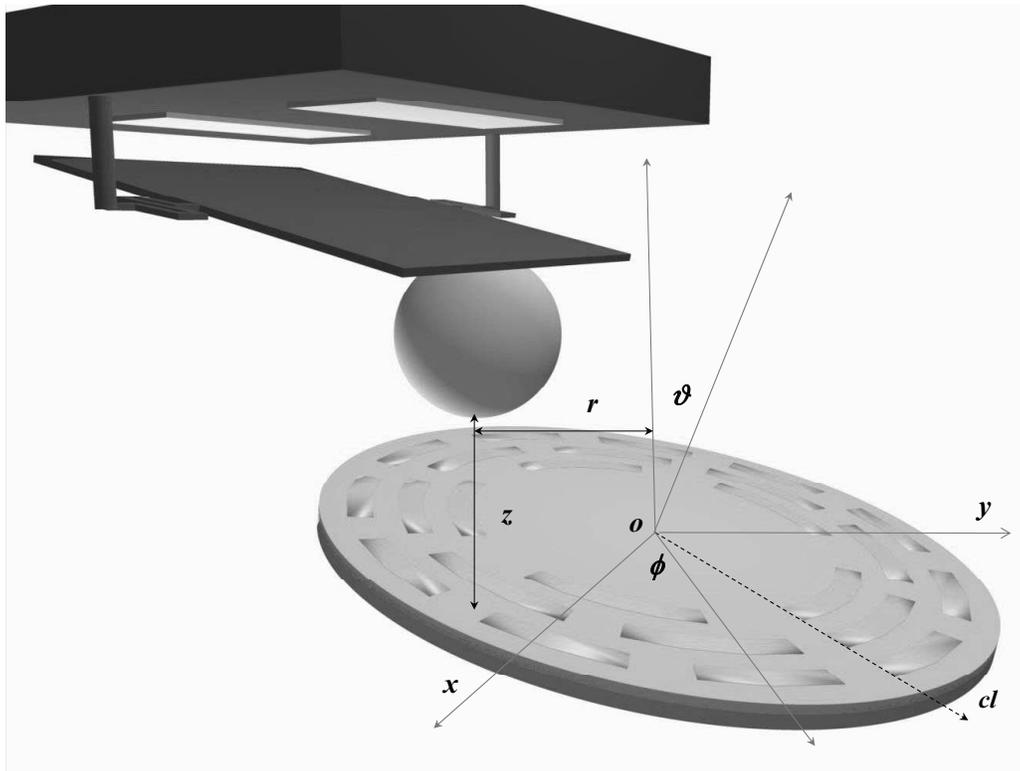}
	\caption{ Schematic of the experimental setup. Three regions with $n_{\rm tr} = 5, 8, 11$ deep trenches are shown.
The actual sample has $n_{\rm tr} = 50, 75, \cdots, 300$ trenches situated at different radii. The region with $n_{\rm tr} = 50$ has inner and outer
 radii of $^{50}r_{\rm i}=4.00$~mm and $^{50}r_{\rm o}=4.15$~mm. A gap of 200~$\mu$m follows.
All gaps have the same radial extent and all trenches have $^nr_{\rm o}- ^nr_{\rm i}=150~\mu$m. The $\{x,y\}$ plane defines
the plane of rotation of the spindle, selected to be parallel to the MTO's substrate. $cl$ is the line where all regions with different
$n_{\rm tr}$ have a common trench interface. $\vartheta$ is the change in the instantaneous axis of rotation, $\phi =\omega t$ is the angle
 of rotation. The distance $z$ is determined from the vertex of the metal-covered sphere to the top of the rotating sample. $r$
 is the distance from the sphere's vertex to the center $o$ of the rotating sample. Displacements $\Delta r$ between $o$ and the
 axis of rotation and the  Au film covering the rotating sample are not shown.}
	\label{Schem}
\end{figure}

The apparatus schematic is shown in Figure~\ref{Schem}. The approach and technique used are
a modification of those described in \cite{17,24}. A metal-coated sapphire sphere is glued
to a high mechanical quality factor $Q$ polysilicon microelectromechanical torsional
oscillator (MTO), which serves as a sensitive force transducer. The oscillator has
a 500~$\mu$m$^2$, 3.5~$\mu$m thick plate anchored to the substrate by two soft,
serpentine-like polysilicon springs \cite{31a}. Underneath the plate two electrodes located
to each side of the
axis of rotation allow to determine the relative motion of the plate with respect to the
substrate by means of the capacitive signal between them and the MTO's plate.

The sample is made of a 1-inch diameter Si wafer where trenches with depth
$d_{\rm T}\approx 50~\mu$m have been made using a deep reactive ion etching approach (DRIE),
based on the Bosch process (a patented process developed by Robert Bosch GmbH \cite{DRIE}).
This process was developed for vertical and deep silicon
etching. Both the sample and the sapphire sphere are subsequently coated with Cr
with $d_{\rm Cr} \approx 10$~nm and a thick enough layer of Au $d_{\rm Au}$ such that from
the point of view of vacuum fluctuations the Au covered bodies  can be considered as
if made from solid Au.  When the sample is rotated in close proximity to the metal coated
sapphire sphere, such that the sphere is alternatively on top of the Au-coated Si wafer or
the deep trench, the sphere experiences a periodic force due to the difference in the
separation-dependent interaction between the sphere and the two heights of the Au-layer
(top of the wafer and bottom of the trench). As the sphere is placed at a position over
$n_{\rm tr}$ alternating Au trenches, the sample's rotational motion (achieved by means of an
air-bearing spindle) is set to the angular frequency
\begin{equation}
\omega=2\pi \frac{f_{\rm r}}{n_{\rm tr}},
\label{omega}
\end{equation}
\noindent
where $f_{\rm r}$ is the operating resonant frequency of the MTO/sphere assembly.
The first harmonic of the force associated with the angular distribution of the
sample will be then naturally selected by the MTO. All other harmonics of the periodic
force and all forces with different angular dependence are outside of the resonance peak
of the MTO and consequently ``filtered'' by the sharp $\delta f \simeq 40$~mHz resonance
peak of the oscillator.

The sphere-MTO system is mounted onto a piezo-driven 3-axis computer controlled flex system (MadCity Labs). The position stability is better than 0.1~nm over 10 hours on all three axes.
The piezo driven stage is mounted on a stepper-motor driven 5 axis stage (Newport).
This stepper-motor stage is used to achieve the initial positioning of the sphere relative to
the rotating sample. Each step on the motor has an amplitude of approximately 9 nm,
translating into a minimal angular deviation of about 0.25~$\mu$rad. After the initial alignment
is achieved with the 5-axis stage, the final linear displacements and positioning are achieved
with the piezo-driven stage. All non-metallic parts close to the MTO are covered with Au-coated mylar or Au-coated Al-foil. Similarly, all Al surfaces (which were observed to produce a drift
of electrostatic nature), are also covered with Au-coated mylar. The mechanical arm between
the rotating sample and the MTO is about 10~cm. The temperature in the chamber is kept at
$T = (295.25 \pm 0.01)$~K, a few degrees above room temperature by means of a standard
combination of heaters and temperature controller (LakeShore). Variations in the controlled
temperature yields observed position drifts of approximately  0.8~nm/hr. The relative drift
between the MTO and the rotating sample is  monitored by continuously measuring the
capacitance between an L-shaped piece attached to the MTO holder and two orthogonal plates
attached to the base of the vacuum chamber \cite{Kolb}.  A two-color interferometer is used
to monitor the $z$ axis separation. Minimum detectable changes $\sim 0.1$~nm along all three
axes are corrected by supplying the appropriate signal to the piezo stage.  The whole
vacuum chamber is mounted into an actively controlled air-damping table (TMC Corporation).
The table and all connections, both electrical and mechanical,  are isolated from vibration
sources  by sand boxes. The combination of vibration isolation systems yields
peak-to-peak vibrations with $z_{\rm pp} < 0.02$~nm
(the detection limit in the accelerometer) for frequencies above 10~Hz. Furthermore, the
active sensing apparatus being a torsional pendulum, it does not effectively couple
to center-of-mass motion associated with vibrations. The high quality factor in the oscillator
is achieved by pumping the system to $P\leqslant 10^{-6}$~Torr (maintained during each run) by
a combination of mechanical, turbomolecular and chemical pumps.

In the air-bearing spindle (KLA-Tencor), the thin air-layer between the rotor and its
encasing makes the system very compliant. On the other hand, the large air flow needed to
operate the spindle required the design and construction of a special seal \cite{17}.

\subsection{Sample Preparation and Characterization}\label{char}

The radius of the sapphire sphere covered with Cr and Au
($d^{\rm sphere}_{\rm Au} \approx 250$~nm) is determined by scanning electron microscopy to
be $R =$(149.7 $\pm$ 0.2)~$\mu$m. The deposited Au on the sapphire sphere is characterized
by atomic force microscopy (AFM) images, and the rms roughness is found to be
$t_{\rm rms} =0.27$~nm. This was obtained by doing 6 non-overlapping $5 \times 5~\mu$m$^2$
(1024 $\times$ 1024 pixels$^2$) scans over the coated sphere near the position of the
sphere closest to the sample.

As aforementioned, trenches in the rotating sample are fabricated by DRIE followed by
the deposition of $d_{\rm Cr} \approx 10$~nm thick layer of Cr followed by
$d^{\rm sample}_{\rm Au} \approx 150$~nm on a 1 inch diameter $100~\mu$m thick [100] oriented
Si wafer. To make the trenches $\approx 3~\mu$m thick photoresist is spun-coated on the Si wafer,
and using conventional photolithographic approach the photoresist is removed from the pattern
with the concentric sectors (where the trenches would be defined). Afterwards,
C$_4$F$_8$ deposition provides the fluorocarbon coating of all surfaces, and SF$_6$
provides the fluorine for isotropic etching. The fluorine does not etch the fluorocarbon
coating, and sputters it by mechanical etching at the bottom, consequently etching the exposed
Si. The cycle forms nanoscallops on the lateral surface, and it is repeated until the desired
depth of the trenches is obtained. During the process each cycle was tuned to remove
about 150~nm of Si. When the desired depth is achieved a final plasma etching step is used
to remove the residual C$_4$F$_8$. The average position of the formed wall is found to be
nearly vertical (the measured angle as determined by SEM inspection in similar structures
is observed to be larger than 89.5~$^{\rm o}$).   Exposed Au surfaces are characterized by white
light interferometry (WLI) and AFM. Both techniques show an optical quality Au film deposited
on top of the Si wafer. The  1024 $\times$ 1024 pixel$^2$ AFM images obtained over different
$10 \times 10~\mu{\rm m}^2$ regions show position-independent $\approx$~10~nm peak-to-peak
roughness with the rms deviations from the mean level of less than  0.4 nm. It was not possible, however, to
determine the quality and overall thickness of the Cr/Au layers deposited on the sidewalls
and bottom of the trenches. The disk is mounted on the air bearing spindle. It is
optically verified that the center of the disk and the axis of rotation of the spindle
coincided
to better than $\Delta r \approx ~5~\mu$m by measuring the gap between the edge of the disk
and the edge of the indentation where the disk sits. The flatness and alignment of the sample
are checked {\it in-situ} using a fiber interferometer (response time 10 ms). The surface of
the sample is perpendicular to the axis of rotation to better than
$z_{\rm o}$~=~20~nm at $^{300}r$ when rotating the disk at $\omega = 2\pi$~rad/s.

\subsection{Oscillators}\label{MTO}

The MTOs are similar to the ones used in previous experiments \cite{8,9,10,11,17,24,31a}. Differently
from some of the previous measurements and as schematically shown in Figure~\ref{Schem} the
metal coated spheres are glued close to the edge  of the plate of the oscillator. Gluing
the Au-coated  spheres at a distance $b = (239 \pm 4)~ \mu$m  from the axis of rotation reduced
the MTO's natural frequency of oscillation from $f_{\rm o} \simeq 700$~Hz to
$f_{\rm r} = (306.45 \pm 0.02)$~Hz. The quality factor was reduced from
$\approx 9000$ to $Q =4850$.

\begin{figure}[!b]
	\centering
	\includegraphics[width=10.5cm]{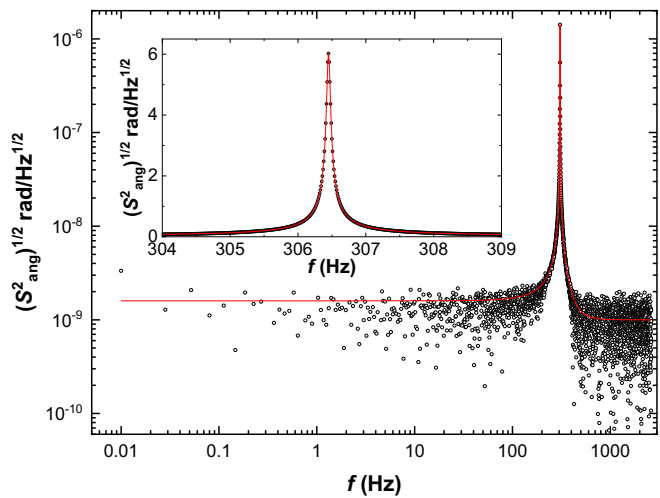}
	\vspace{-0.3cm}
	\caption{Free-standing frequency response of the oscillator with the Au-coated sphere glued
to it. The high frequency response shows the limits of the detection circuit.  The inset shows
an expanded view of the resonance with an average of 100 different spectra. The red solid line
is a fit using (\ref{power}) with a detection flat spectral density noise of
$1.2 \times 10^{-9}~{\rm rad}/\sqrt{\rm Hz}$. Points below 0.025~Hz where $1/f$-noise is
measurable have been excluded from the fit.}
	\label{osc}
\end{figure}
The power spectral density $S^2_{\alpha}(f)$ of the oscillator is shown in Figure~\ref{osc}.
For a torsional  simple harmonic damped oscillator driven by thermal fluctuations the angular response in the torsional angle $\alpha$ is \cite{Casimirbook}
\begin{equation}
S^2_{\alpha} (f)= \frac{2k_{\rm B}T}{\pi\kappa Q f_{\rm r}}\frac{f_{\rm r}^4}{(f_{\rm r}^2-f^2)^2+f^2f_{\rm r}^2/Q^2}+ S^2_{\rm elec}\, ,
\label{power}
\end{equation}
\noindent
where an independently determined  flat detection noise term $S^2_{\rm elec}$ associated with
the electronic measurement setup \cite{17} has been added. In (\ref{power}), $k_{\rm B}$
is Boltzmann's constant, $T$ the temperature at which the experiment is performed, $\kappa_{\rm MTO}$ is
the MTO's torsional constant. Doing the measurement at resonance, where the $1/f$ term and
the detection noise are negligible, it is found that the minimum detectable force
(per Hz$^{1/2}$) is
\begin{equation}
\frac{1}{b}\sqrt{\frac{2 \kappa_{\rm MTO} k_{\rm B} T}{\pi Q f_{\rm r}}} \sim 6
\frac{\rm fN}{\sqrt{\rm Hz}}.
\label{Fmin}
\end{equation}
\noindent
With the achieved temperature control, the drift in the resonant $f_{\rm r}$ is less
than 0.3~mHz/hr under operating conditions.

\subsection{Electrostatic Calibration and Separation Determination}\label{sep}

The system's calibration is performed similarly to what was done in \cite{elcal}.
An optical fiber is rigidly attached to the MTO-sphere assembly, and a two-color interferometer
is used to measure the distance between the end of the fiber and a stationary platform.
Simultaneously $f_{\rm r}$ and the angular deviation of the MTO are recorded as the sphere is
moved closer to the sample. From the change in $f_{\rm r}(z)$ the gradient of the
interaction between the sphere and the plate can be obtained when a potential difference is
applied between them. Comparing the separation dependence of the gradient of the interaction
with that of the known sphere-plate electrostatic interaction
\begin{eqnarray}
&&
F_e(z,V)= -2\pi \varepsilon_{0} (V - V_{\rm o})^2 \sum_{n=0}^\infty \frac{\coth(u) - n \coth(nu)}{\sinh(nu)}
\nonumber \\
&&
\phantom{F_e(z,V)}
 = -2\pi \varepsilon_{0} (V - V_{\rm o})^2 \sum_{m=0}^7A_m q^{m-1},
\label{el2}
\end{eqnarray}
\noindent
the parameters of the system are obtained. In (\ref{el2}) $\varepsilon_{0}$  is
the permittivity of free space (in SI units), $V$ is a potential applied to the sample
(the sphere-oscillator assembly is always kept grounded) and $V_{\rm o}$ is a residual
potential difference between the plate and the sphere, $u = 1+z/R$, $A_m$ are fitting
coefficients, and $q = z/R$. While the full expression is exact, the series is slowly
convergent, and it is easier to use the shown approximation developed in \cite{fit}.
Using this approach, $\kappa_{\rm MTO} = (1.07 \pm 0.01)\times 10^{-9}$~Nm/rad is obtained, with
$V_{\rm o}$ of the order of a few mV. For all configurations used,
$V_{\rm o}$ was checked to be position and time independent. As customary in these experiments,
the differential measurements are performed with $V=V_{\rm o}$ to minimize
the electrostatic contribution.

In order to simplify the data acquisition and control of the system, during the experiment
the two-color interferometer is used such that it controls the separation between the
sphere-MTO assembly and a fixed platform, yielding a measurement of the distance
$z_{\rm meas}$ between the end of the fiber and the fixed platform.  In order to find
the separation $z$ between the sphere and the top of the rotating disc
\begin{equation}
z=z_{\rm meas} -D_1-D_2-b\alpha,
\label{dist}
\end{equation}
\noindent
the quantity $D_1+D_2$ is obtained from the electrostatic calibration. In (\ref{dist}),
$D_1$ is the fixed distance between the end of the fiber and the vertex of the sphere when
the system is relaxed, and $D_2$ is the distance between the fixed platform and the top of
the rotating sample. At each measurement position the torsional angle $\alpha$ is measured with
the sphere on top of the Au region, away from the regions with trenches (i.e. no signal is
expected at $f_{\rm }$ in this situation) and the value of $\alpha$ (which is always smaller
than $10^{-5}$~rad) is determined from the difference in capacitance between the MTO's plate and
the two underlying electrodes. Hence, from the measured value $z_{\rm meas}$, the fitted
$D_1+D_2$, the optically determined $b$ and the capacitively determined $\alpha$, the
separation $z$ is found.

\subsection{Results}\label{res}

\begin{figure}[!h]
	\centering
	\includegraphics[width=.7\columnwidth]{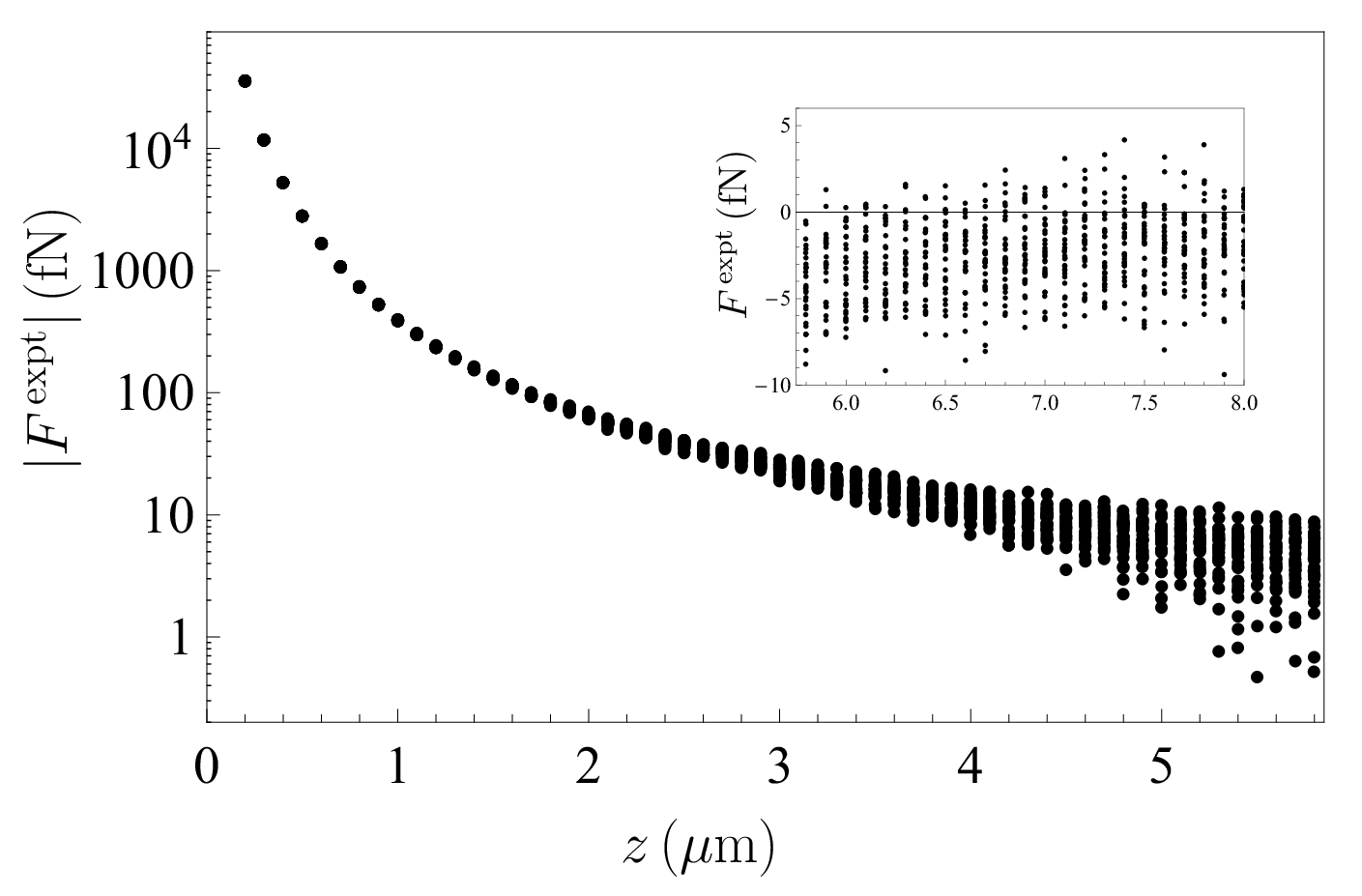}
	\caption{All measured magnitudes of the Casimir force between an Au-coated sphere and an Au-coated
plate obtained from the first harmonic of the interaction as the plate rotates under the
sphere are shown as a function of separation (in a logarithmic scale). The inset shows
the force values over the region of large
separations (in a linear scale).}
	\label{results}
\end{figure}

Extraction of the data is done assuming the interaction between the sphere and the trenches
follows a Heaviside function as the sample rotates under the sphere. Under this condition,
the force values reported are $\pi/4$ times the measured value at the corresponding harmonic.
At each point the force value was determined from the first harmonic.
The force measurements were repeated 30 times in the separation region from 200~nm to
$8~\mu$m with a step of 100~nm.
Each repetition was measured with an integration time of 100~s.
In Figure~\ref{results} the obtained results (30 force magnitudes at each separation) are
shown in logarithmic scale over the region from 0.2 to $5.8~\mu$m.
In the inset, the measured forces over the region from 5.8 to $8~\mu$m are presented
in linear scale. The form of the distribution law and the random errors of the
measurement results are considered in Section~6.1.

\section{Systematic Errors and Edge Effects}\label{SE}

In this section, we consider different contributions to the total systematic error in
measuring the Casimir force by means of a micromechanical torsional oscillator. Taking into
account that in our case the disc in Figure~\ref{Schem} is not flat but contains deep,
concentrically situated trenches, we also investigate the size of possible errors in the
measured force which could arise from edge effects.

\subsection{Contributions to the Systematic Error}

The success of the experiment resides in having a controlled metrological environment for
the interaction and separation (as described in Section~\ref{sep}) but since also a
lock-in amplifier technique is used, it is imperative to preserve a tight time and
frequency synchronization. Time synchronization is achieved by focusing a diffraction limited
laser on the rotating sample at a distance $r\approx 9$~mm from its rotational axis. The sample
itself has a region where no Au is deposited located in $r\in[8.5,9.5]$~mm subtending an angle
of $2\times10^{-4}$~rad. The leading edge of this sector is along the {\it cl}-line.
As the sample rotates, the change in reflection of the focused laser is used as a trigger
for all timed events. It has been verified that this trigger  lags by
$\tau_{\rm lag} = 10^{-6}/f$. The rotation frequency is obtained
by finding the maximum of the thermally induced peak shown in Figure~\ref{osc} with an
accumulation time of 100 s. The required multiple of this signal is synthesized and fed to
the air bearing spindle.

In general, with the sphere placed at $^{300}r_{\rm i} +75~\mu$m, the air bearing spindle
was rotated at $\omega=2\pi f_{\rm r}/300$. In this manner, a force arising from the difference
in the Casimir force between the metal coated sphere and the layered structure manifests itself
at $f_{\rm r}$ even though  there are no parts moving at $f_r$. Using lock-in detection at
$f_{\rm r}$ signals which are small but could show in conventional experiments are removed by
the averaging provided by the rotating sample and the high-$Q$ of the MTO. While this approach
is employed to obtain the interaction, the large range of separations used and the consequent
large change in the strength of the interaction presents a drawback: at the short end of the
range in $z$ the relatively large force difference between the situations when the sphere is
on top of the ridges or trenches would cause the oscillator to behave non-linearly or break.
To prevent this, the measurements at small $z$ are detuned from $f_{\rm r}$. Since now the
system is at a steep part of the resonance curve, the errors in frequency stability are
amplified when compared to the corresponding errors at large $z$ when the measurements can
be done at resonance.

It is known from previous studies \cite{17,24} that the air bearing spindle has  a
revolution impulsive kick on the order of $\vartheta \sim 10^{-7}$~rad. Fortunately in
the measurements performed in this study this does not affect the measurement.
The system was positioned such that when the impulsive kick happens, the sphere is located
over a trench and there is no effect on the measurement. In all measurements it is observed
that when $z > 20~\mu$m the expression shown in (\ref{Fmin}) is verified: As the integration
time $\tau$ increases, the detectable force decreases as 5.8~fN/$\sqrt{{\rm Hz}\,\tau}$.

\begin{figure}[!t]
	\centering
	\includegraphics[width=.7\columnwidth]{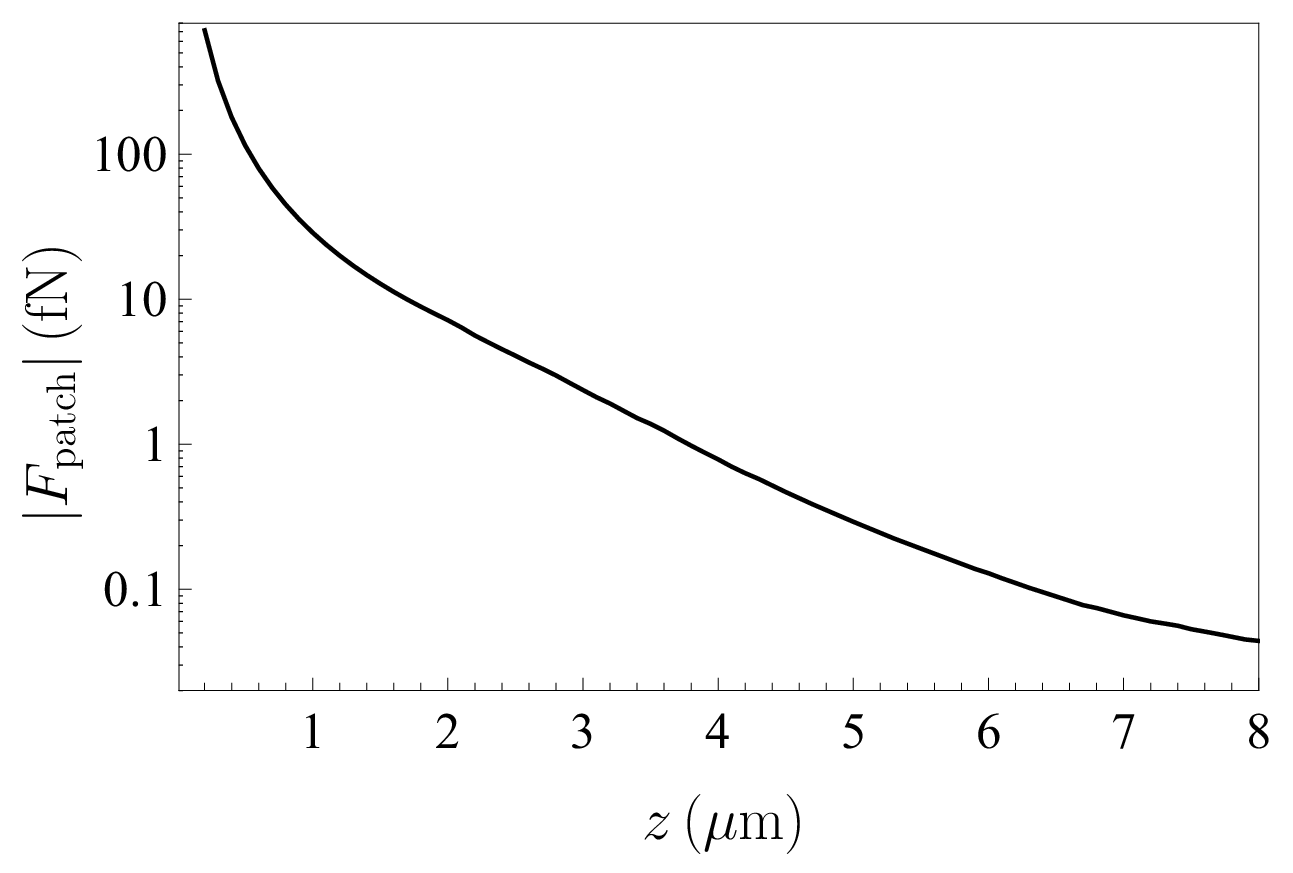}
	\caption{An estimated magnitude of the force originating from
surface patches as a function of separation (in a logarithmic scale). }
	\label{figPatch}
\end{figure}

Contributions to the systematic errors of measurements are summarized in Table\ref{tableI}.
\begin{table}[H]
\caption{Systematic errors in the separation and determination of the measured force.
The errors in $b$ ($4~\mu$m) and $\alpha$ (0.8~nrad) do not contribute to the overall error
of separation. The listed error in $D_1+D_2$ is obtained from the electrostatic calibration.
The error reported in $z_{\rm meas}$ is larger than the local measured error of 0.2 nm
due to the fact that the rotating sample has measured height difference of $\approx $1.2 nm.
This is not adjusted during rotation. The force measurement error varies from 85~fN at the
shortest separation to 0.5 fN at $z \geqslant 1 \mu$m. All errors are at the 95\% confidence level.\label{tableI}}
\begin{tabular}{ccccccc}
\toprule
	& $D_1+D_2$~[nm]	& $z_{\rm meas}$~[nm] & Flatness of wafer~[nm] \\
\midrule
{\bf Separation}	& 0.6 	& 0.2 &	1.2	\\
\midrule
\midrule
&Detection~[fN]&Calibration~[fN] & Measurement~[fN] \\
\midrule
{\bf Force}  &0.6 &0.2 &[85, 0.5]  \\

\bottomrule
\end{tabular}
\end{table}
There is one more uncertainty in an interpretation of the measured force signal as the
Casimir force. It is connected with the contribution of patch potentials.
The sample was characterized by Kelvin probe microscopy, and the potential distribution was
found to be similar to the one in \cite{KP} with $V_{\rm rms}\approx 12~$mV and average size of
patches $\bar{l}\approx 250~$nm. At the smallest separation $z_1=200~$nm, the attractive force
due to the presence of patches can be estimated as $|F_{\rm patch}|\approx 0.7~$pN
(see Figure~4 of \cite{Beh12}), i.e., of about 2\% of the measured Casimir force.
Using the asymptotic expression \cite{Beh12}
\begin{equation}
F_{\rm patch}(z)=-\pi\zeta(3)R\frac{\varepsilon_0 V_{\rm rms}^2\bar{l}^2}{2z^3},
\label{eqPatch}
\end{equation}
\noindent
which is well applicable at $z>7~\mu$m, the estimated magnitude of the force originating from the
surface patches is shown in Figure~\ref{figPatch} over the entire measurement range.

\subsection{Investigation of Edge Effects}

In what follows we analyze the edge effects, i.e. possible deviations of a signal shape
from the Heaviside function during rotation. In order to do this analysis, 21 harmonics
of the measured signal were determined at four separations $z =$ 200, 500, 1000, and 5000~nm.
At these separations all measured force values are negative (see Figure~\ref{results}).
Below we consider the force magnitudes.
To measure the $m$-th harmonic of the signal, the sample was rotated at an angular
frequency $\omega_{m} =2\pi f_{\rm r}/(m n_{\rm tr})$ (recall that $n_{\rm tr}$ is the number of alternating
Au trenches).
Furthermore, the even and odd contributions to the signal are selected by setting the phase of the lock-in detection to 0
when the sphere crosses the line {\em cl} (see Figure~\ref{Schem}).

The general properties of the Fourier series of the signal are what follows next.
Let $\hat{\phi}$ be an angle variable, defined such that  $\hat{\phi}$ changes by $2 \pi$ as
one traverses a Au sector and a trench. Up to random patch effects and other sources of noise,
the signal $F(\hat{\phi})$ is then $2 \pi$ periodic:
 \begin{equation}
F(\hat{\phi}+ 2 \pi)=F(\hat{\phi})\;.\label{per}
 \end{equation}
 \noindent
If the origin of the angles is taken to be in the center of a Au sector, the force
signal is also expected to be symmetric with respect to an inversion of  $\hat{\phi}$:
\begin{equation}
F(-\hat{\phi})=F(\hat{\phi})\;.\label{inv}
 \end{equation}
 \noindent
It follows from (\ref{per})  and (\ref{inv}) that the Fourier expansion of the signal  contains only cosines and, thus, is of the form:
 \begin{equation}
F(\hat{\phi})=\frac{a_0}{2}+\sum_{m=1}^{\infty} a_m \cos(m \hat{\phi})\;.\label{fou1}
 \end{equation}

To proceed, we make the assumption that the signal can be represented as a  sum of the
step function $F\; \chi(\hat{\phi})$ plus an edge correction $f(\hat{\phi})$:
 \begin{equation}
F(\hat{\phi})=|F|\;\chi(\hat{\phi}) + f(\hat{\phi})\;,\label{signal}
 \end{equation}
\noindent
where $F$ is the  force between an Au sphere and a {\it homogeneous} Au plate,
and $\chi(\hat{\phi})$ is the $2\pi$ periodically continued step function of the
interval $[-\pi, \pi]$, which is equal to unity for
$-\pi/2 \leqslant \hat{\phi} \leqslant \pi/2$ and zero elsewhere. The correction
$ f(\hat{\phi})$ represents the effect of the edges, which includes both edge-corrections
to the Casimir force as well as stray electrostatic forces arising from charges localized
along the edges.  We assume that  $ f(\hat{\phi})$ is {\it localized} in a narrow region
of angular width $\epsilon$ across the edges of the Au sectors.  According to (\ref{signal})
we decompose the Fourier coefficients $a_m$ as:
 \begin{equation}
a_m= a_m^{(\rm step)}+\delta a_m\;,
 \end{equation}
\noindent
where $a_m^{(\rm step)}$ is the contribution of the step function and $\delta a_m$ is the edge correction. By a straightforward computation one finds that the coefficients $a_m^{(\rm step)}$ are zero for even $m$, while for odd $m$ one finds:
 \begin{equation}
a_{2p-1}^{(\rm step)}=-(-1)^p \frac{2 |F| }{\pi (2p-1)}\;,\;\;\;p=1,2,\dots \label{stepfou}
 \end{equation}

On the other hand, for the corrections $\delta a_m$ one finds:
 \begin{equation}
\delta a_m=\frac{2}{\pi} \int_{\pi/2-\epsilon}^{\pi/2+\epsilon} d \hat{\phi}\;  f(\hat{\phi})\;\cos(m\hat{\phi})=\frac{2}{\pi} \int_{-\epsilon}^{\epsilon} d x\;  f\left(\frac{\pi}{2}+x\right)\;\cos\left[m\left(\frac{\pi}{2}+x\right)\right]\;.
 \end{equation}
Since the width $\epsilon$ of the edge region is small, for $m$ not too large it is possible
to take the Taylor expansion of the cosines in power of $x$.  By Taylor expanding the
cosines around $\hat{\phi}=\pi/2$ up to order $x^2$ included, for even $m=2p$, $p=1,2,\dots$
one finds:
 \begin{equation}
\delta a_{2 p}= \frac{2}{\pi} (-1)^p \left[f^{(0)}- 2 p^2 f^{(2)}\right]\;,\label{delaev}
 \end{equation}
\noindent
while for odd $m=2p-1$, $p=1,2,\dots$ one finds:
 \begin{equation}
\delta a_{2 p-1}= \frac{2}{\pi} (-1)^p (2p-1) f^{(1)}\;,\label{delaodd}
 \end{equation}
\noindent
where $f^{(q)}$ is the $q$-th moment of the edge correction:
 \begin{equation}
 f^{(q)}=\int_{-\epsilon}^{\epsilon} d x\; x^q\; f\left(\frac{\pi}{2}+x\right).
 \end{equation}
\noindent
Combining (\ref{stepfou})--(\ref{delaodd}), we find that the leading Fourier coefficients
in (\ref{fou1})  are those with odd $m=2p-1$:
 \begin{equation}
a_{2 p-1}=-\frac{2}{\pi}  (-1)^p \left[\frac{ |F| }{ 2p-1}-(2p-1) f^{(1)}\right]\;,
 \end{equation}
\noindent
while for even $m=2 p$ the Fourier coefficients coincide with the edge correction:
 \begin{equation}
a_{2 p}=\frac{2}{\pi} (-1)^p \left[f^{(0)}- 2 p^2 f^{(2)}\right]\;.
 \end{equation}

In the experiment, angles are measured starting from the edge of a Au sector. Thus we define
the shifted angle variable $\bar{\phi}$ such that $\hat{\phi}=\bar{\phi}-\pi/2$. When
re-expressed in terms of $\bar \phi$, the Fourier expansion in (\ref{fou1}) takes the form:
 \begin{equation}
F(\bar{\phi})=\frac{a_0}{2}+\sum_{p=1}^{\infty} (-1)^p \left\{a_{2 p} \cos(2 \,p \,\bar{\phi})-a_{2p-1} \sin[(2p-1)\bar \phi] \right\}\;.\label{fou2}
 \end{equation}
In reality, the origin of the angles  cannot exactly coincide with the edge of Au sector,
and we should allow for a possible small phase $\delta$. From the sample design and the
time synchronization procedure,  $\delta$ is expected to be of order:
 \begin{equation}
|\delta| \sim 10^{-4}\;{\rm rad}\,.\label{expsh}
 \end{equation}

We define the final angle variable $\phi$ such that $\bar{\phi}=\phi-\delta$. When
the series in (\ref{fou2})  is re-expressed in terms of $\phi$, and only leading terms
in the small phase $\delta$ are retained, one obtains:
\begin{eqnarray}
&&
F(\phi)=\frac{a_0}{2}+ \frac{2}{\pi} \sum_{p=1}^{\infty} \left\{
\vphantom{\left[\frac{|F|}{2 p-1}-(2 p-1)   f^{(1)}\right]}
\left[f^{(0)}-2p^2f^{(2)}\right] \cos(2\, p\, \phi)- |F|\, \delta \cos[(2 p-1)\, \phi] \right.
\nonumber\\
&&
\left.+
\left[\frac{|F|}{2 p-1}-(2 p-1)   f^{(1)}\right] \sin[(2 p-1)\phi]+  f^{(0)} 2\, p\, \delta \sin(2 \,p\, \phi)\right\}\;.\label{four}
 \end{eqnarray}
\noindent
We remind that  the expressions of the Fourier coefficients in the above equation are
valid for harmonics such that $p\,\epsilon \ll 1$ and $p\,\delta \ll 1$.
We assume that both conditions are satisfied for the measured  harmonics. The final
Fourier expansion (\ref{four}) is of the form:
 \begin{equation}
F(\phi)=\frac{a_0}{2}+\sum_{m=1}^{\infty} \left[b_m \sin(m \phi)+c_m \cos(m \phi) \right]\;.
 \end{equation}
\noindent
Its general features are as follows:

1) The dominant terms are sines with odd $m$ and the coefficients:
 \begin{equation}
b_{2 p-1}=\frac{2}{\pi}\left[\frac{|F|}{2 p-1}-(2 p-1)   f^{(1)}\right] \;.\label{sodd}
 \end{equation}

2) The coefficients $b_{2 p}$ of sines with even $m$ are proportional to the phase shift
$\delta$ and to the first moment $  f^{(0)}$ of the edge correction. Moreover, they are linear in the Fourier index $m=2 p$:
 \begin{equation}
b_{2 p}=\frac{2}{\pi} f^{(0)} \, 2\, p\, \delta\;.\label{ssin}
 \end{equation}

3) The coefficients $c_{2 p}$ of the cosines  with even $m$ depend on the
moments $  f^{(0)}$ and $  f^{(2)}$ of the edge correction,  and have a quadratic
dependence on the index $m=2 p$:
 \begin{equation}
c_{2 p}=\frac{2}{\pi}  \left[f^{(0)}- 2 p^2 f^{(2)}\right]\;.\label{cc2p}
 \end{equation}

Finally, the coefficients of the cosines with odd $m$ are proportional to the phase
shift $\delta$ and are independent of the order $m=2p-1$:
 \begin{equation}
c_{2 p -1}=-\frac{2}{\pi}  \, |F|\, \delta\;.\label{ccodd}
 \end{equation}

Based on the above equations, one can predict that the Fourier coefficients satisfy
the following hierarchy:
 \begin{equation}
b_{2 p-1} \gg c_{2 p} \gg b_{2 p} \sim c_{2 p -1}\;,
 \end{equation}
\noindent
which was verified to be true.

\begin{figure}[t]
\hspace*{1.5cm}
\includegraphics[width=.7\columnwidth]{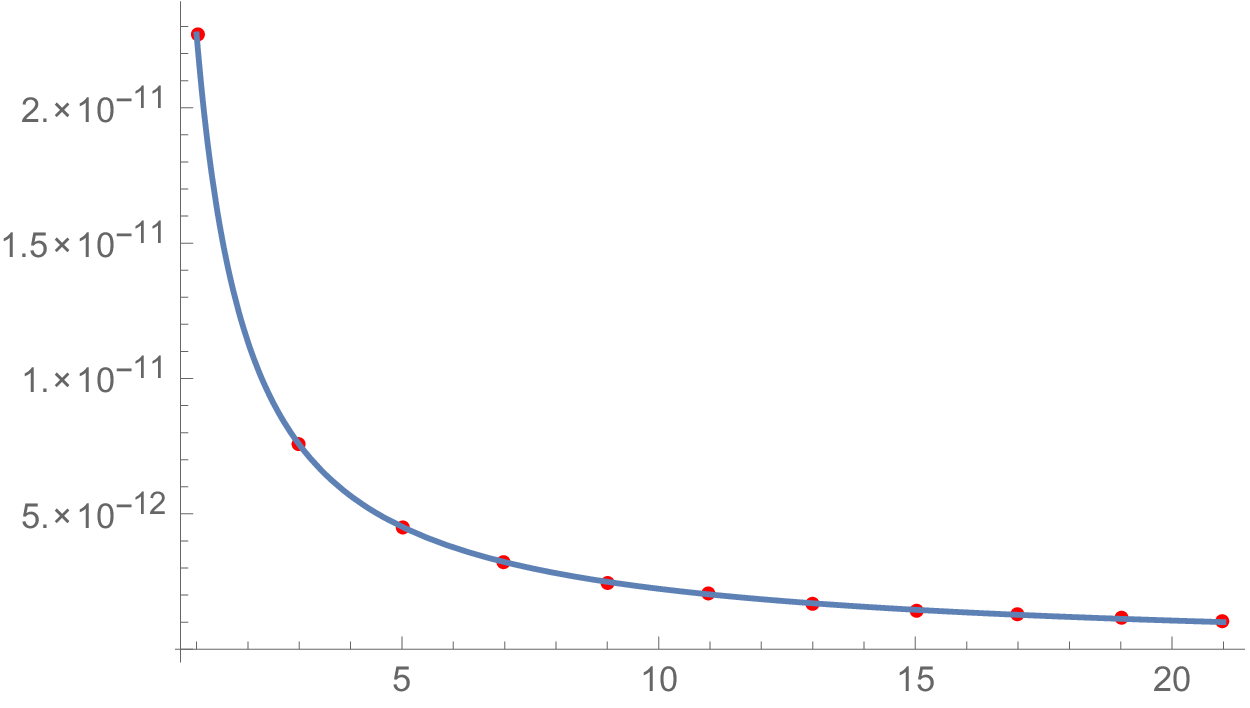}
\caption{\label{oddsines}  Fourier coefficients $b_m$ (in N) for sines
 with odd $m =1,3,5,\dots$ at $z=200$ nm. The solid line is a best fit of the data
 by the curve $b_m = u_1/m + u_2\, m$.  }
\end{figure}
\begin{figure}[t]
\hspace*{1.5cm}
\includegraphics[width=.7\columnwidth]{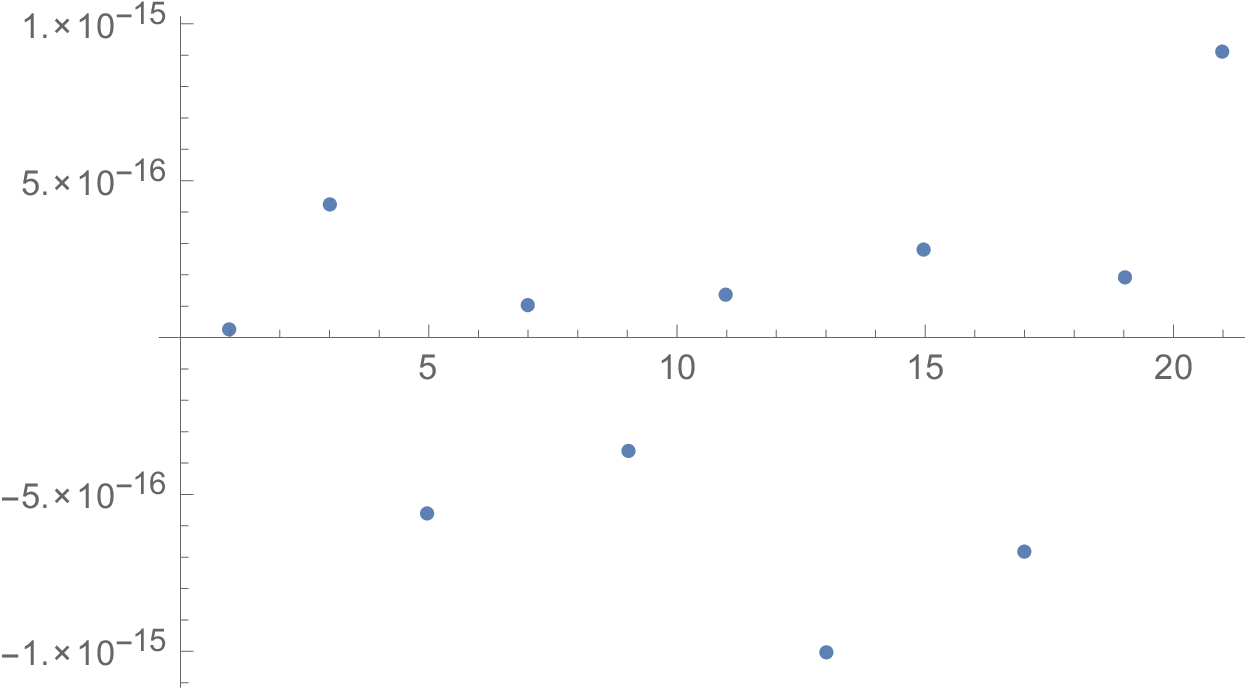}
\caption{\label{residualsoddsines} Differences (in N) between the measured Fourier coefficients $b_m$ from Figure~\ref{oddsines}  and the best fit  curve $b_m$.  }
\end{figure}
We used  (\ref{four})  to  analyze the 21 measured harmonics for the nominal  sphere-plate separations $z=200$, 500, 1000, 5000~nm.
As an example, below we present the results obtained at $z=200~$nm.
In Figure~\ref{oddsines} we show the data for the sines with odd $m$ for $z=200$ nm, and the fit by a curve of the form  $b_m= u_1/m + u_2\, m$. The agreement is excellent as it can be seen from Figure~ \ref{residualsoddsines} where we show the differences between the data and the fitting curve. Since for each measured harmonic  only one measurement was made, it was in principle impossible to determine their error. To circumvent this problem, we assumed that the error for the 21 harmonics is the same, and thus we estimated the common standard deviation $\sigma$ of the odd coefficients $b_{2 p-1}$   by the formula:
 \begin{equation}
\sigma_{\rm 200 nm}^2=\frac{1}{9} \sum_{p=1}^{11}
\left[b_{2p-1}- \frac{\bar{u}_1}{2p-1} - \bar{u}_2\,(2p-1)\right]^2\;,
 \end{equation}
where $\bar{u}_1$ and $\bar{u}_2$ are best fit values.  The sum over $p$ is divided by 9 because there are 11 Fourier coefficients and two free parameters ($u_1$ and $u_2$). We obtained:
 \begin{equation}
\sigma_{200 \rm nm}=0.6\;{\rm fN}\;.
\label{eq28}
 \end{equation}
\noindent
The values of $\sigma$  that were obtained for the other Fourier coefficients $b_{2p }$, $c_{2 p-1}$ and $c_{2p }$ have a similar magnitude, for all the four considered separations. For example, from $b_{2p-1}$ we find:
 \begin{equation}
\sigma_{500 \rm nm}=0.5\;{\rm fN}\;,\quad
\sigma_{1000 \rm nm}=0.4\;{\rm fN}\;,\quad
\sigma_{5000 \rm nm}=0.6\;{\rm fN}\;.
\label{eq29}
 \end{equation}
\noindent
Using the estimates (\ref{eq28}) and (\ref{eq29}) of the standard deviations, we were able to determine the forces and edge corrections at the desirable confidence level.

The best fit of the Fourier coefficients $b_m$, with  odd $m$ provides the following estimate of the force at $z=200~$nm:
 \begin{equation}
F^{\rm expt} =(-3.5675  \pm 0.0003  ) \times 10^{-11}\;{\rm N}  \label{estfor}
 \end{equation}
and of the edge correction
 \begin{equation}
 f^{(1)}=(6.01   \pm 0.06)\times 10^{-15}\;{\rm  N}\;, \label{f1200}
 \end{equation}
where both errors correspond to the 99\% confidence level.

\begin{figure}[t]
\hspace*{1.5cm}
\includegraphics[width=.7\columnwidth]{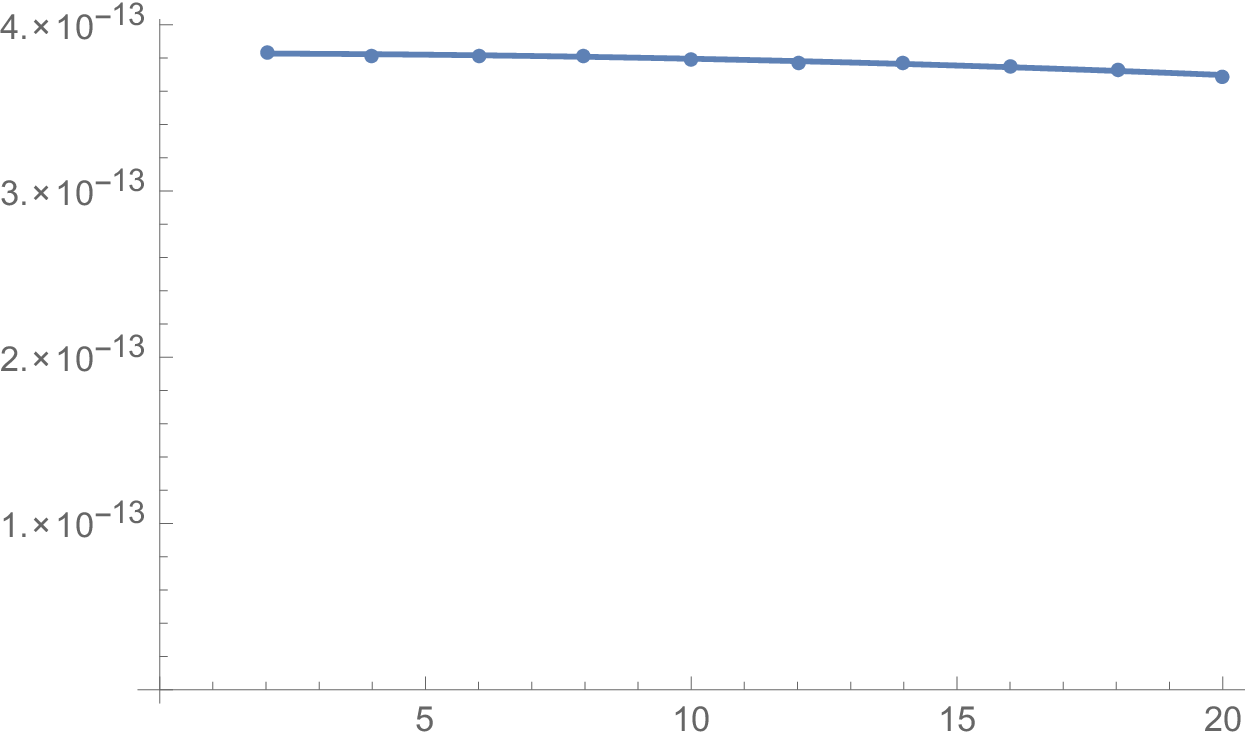}
\caption{\label{evencosines}  Fourier coefficients $c_m$ (in N)  for cosines
 with even $m =2,4,6,\dots$ at $z=200$ nm. The solid line is a best fit of the data
 by the curve $c_m = v_0+v_2\,m^2$.  }
\end{figure}
\begin{figure}[t]
\hspace*{1.5cm}
\includegraphics[width=.7\columnwidth]{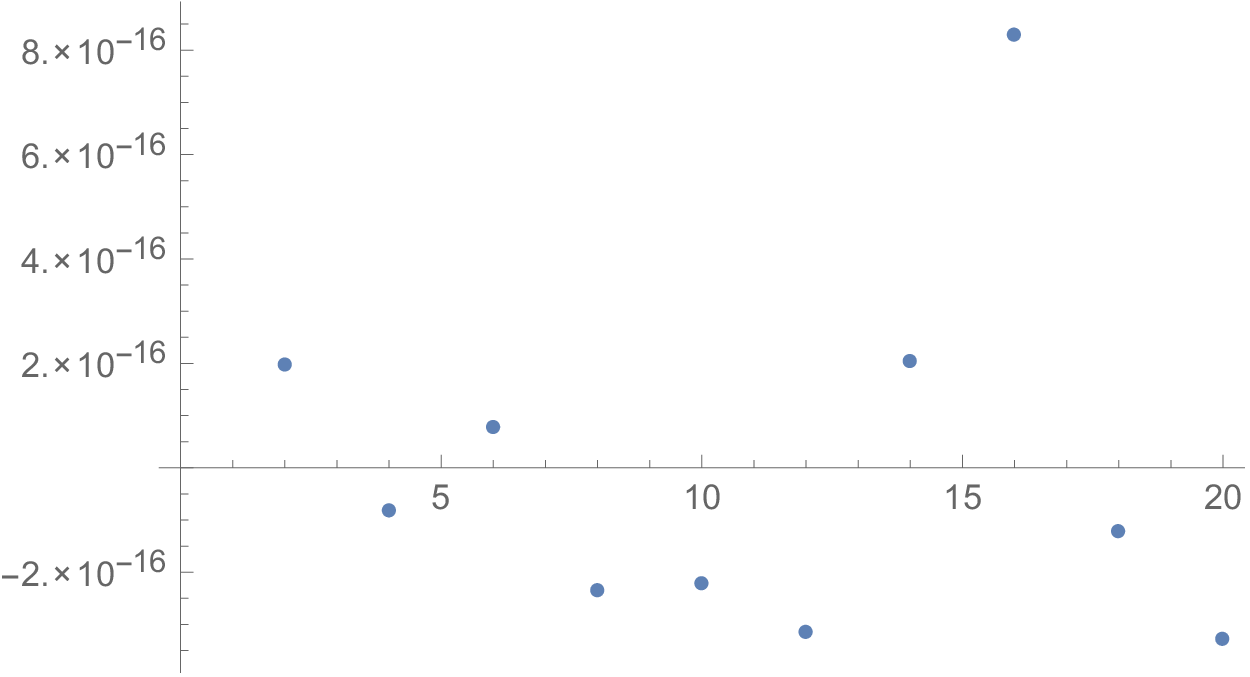}
\caption{\label{residualsevencosines} Differences (in N) between the measured Fourier coefficients $c_m$ from Figure~\ref{evencosines}  and the best fit  curve $c_m$.  }
\end{figure}
Next we consider the data for cosines with even $m$.
These are shown in Figure~\ref{evencosines}.  The solid line is a fit with a quadratic
curve $c_m=v_0+v_2 m^2$, as per (\ref{cc2p}). The agreement is excellent as it can be seen
from Figure~ \ref{residualsevencosines} where we show the differences between the data and
the fitting curve.

{}From the fit we obtain:
 \begin{equation}
 f^{(0)}=(6.013 \pm 0.010) \times 10^{-13}~\mbox{N}, \quad
 f^{(2)}=(1.03 \pm 0.10) \times 10^{-16}~\mbox{N}, \label{f2200}
 \end{equation}
where both errors correspond to the 99\% confidence level. Equations~(\ref{f1200}) and
(\ref{f2200}) show that there indeed is an edge effect, but fortunately it is
negligibly small.

\begin{figure}[t]
\hspace*{0cm}
\includegraphics[width=.5\columnwidth]{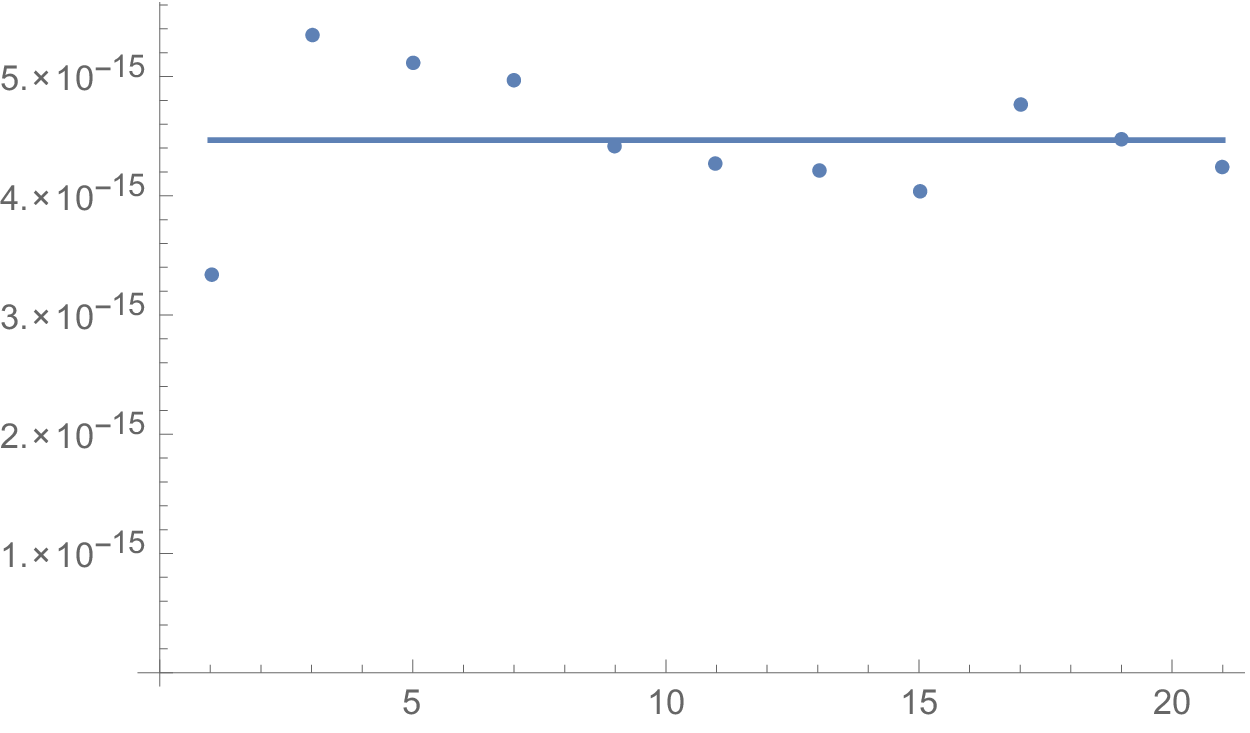}
\vspace*{2mm}
\includegraphics[width=.5\columnwidth]{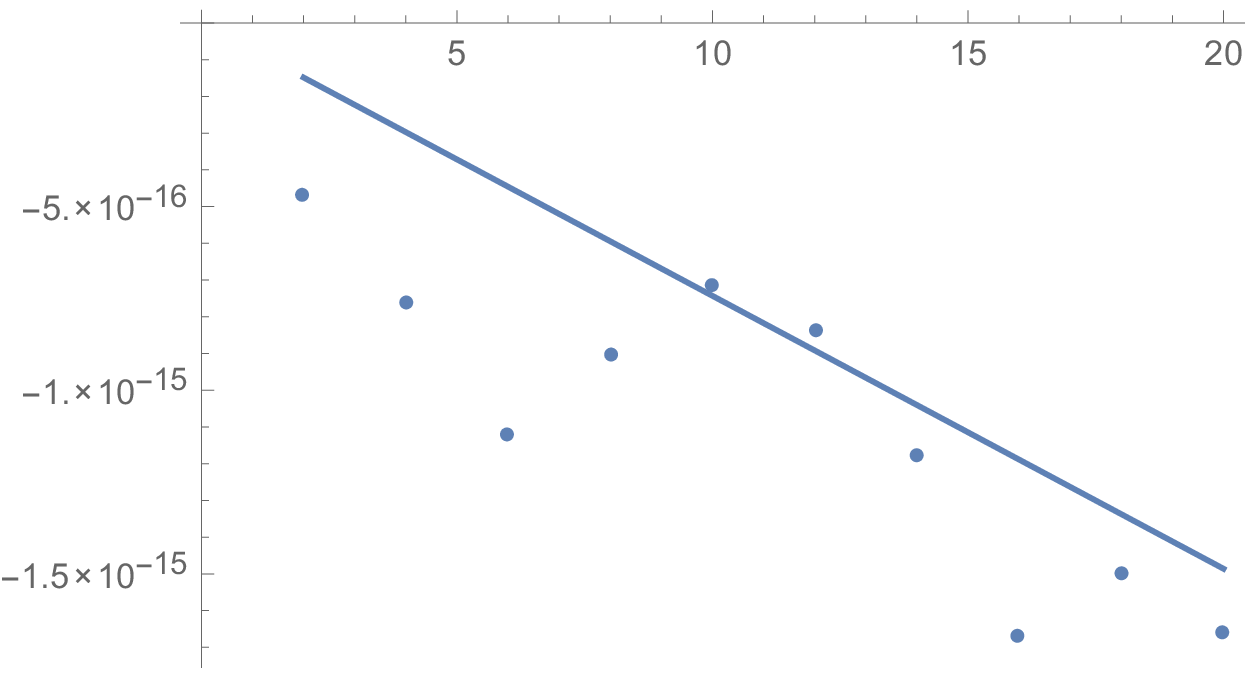}
\caption{\label{oddcosines}  Fourier coefficients in N   at $z=200$~nm
(left) $c_m$ for cosines with odd
$m =1,3,5,\dots\,$, where the line is the best constant fit, and
(right)  $b_{m}$   for sines with
 even $m=2,4,6,\dots$, where the solid line is obtained using (\ref{ssin}) with previously determined $\delta$ and $f^{(0)}$ values.}
\end{figure}

Now we turn our attention to the data for cosines with odd $m$, which, according
to (\ref{ccodd}), should be independent of the Fourier index $m=2p-1$.
The respective data are shown in Figure~\ref{oddcosines} (left).
As is seen in this figure, the data are  really almost independent of the Fourier index.
The observed deviations have the character of statistical fluctuations.  The fit gives:
\begin{equation}
\delta=-2 \times 10^{-4}\; {\rm rad}
 \end{equation}
in accordance with the expected magnitude of $\delta$ in (\ref{expsh}).

We finally consider the Fourier coefficients for sines with even $m$. The data are shown
in Figure~\ref{oddcosines} (right), together with a plot of the  linear dependence (\ref{ssin}).
The line drawn in Figure~\ref{oddcosines} (right) uses the values of $\delta$ and $f^{(0)}$
determined previously. The fairly good agreement between (\ref{ssin}) and the data provides
a further validation of the theoretical model of the signal (\ref{four}).

The overall analysis of the signal rules out any important influence of edge effects and justifies modeling the interaction between the sphere and the sample by a Heaviside function.

\section{Exact Evaluation of the Casimir Force in Sphere-Plate Geometry Using the
Scattering Formula}

In this section, we calculate the Casimir force between an Au sphere and an Au plate
on the basis of first principles of quantum electrodynamics. This is especially important
for subsequent comparison between experiment and theory at separations of a few micrometers
where the approximate methods like the proximity force approximation (PFA) become less exact. Below we
review the theoretical basis for the numerical
determination of the Casimir force using electromagnetic plane waves which has been
employed to determine the exact results presented in this paper. We aim at a coherent
presentation of results presented in different places in the literature
\cite{26,27,28}.

Within the scattering formalism, the Casimir free energy $\mathcal{F}$ between a sphere and a plane at a surface-to-surface distance $z$ can be expressed through
\cite{40a,40b,40c,Lambrecht2006,Rahi2009}
\begin{equation}\label{eq:calF}
\mathcal{F} = k_B T \sum_{l=0}^\infty{}' \log\det[1-\mathcal{M}(i\xi_l)],
\end{equation}
where the primed sum indicates that the $l=0$ term is weighted by a factor of $1/2$, the Matsubara frequencies are given by $\xi_l=2\pi k_B T l/\hbar$ and the round-trip operator is defined as
\begin{equation}\label{eq:round-trip_operator}
\mathcal{M} = \mathcal{T}_\mathrm{PS}\mathcal{R}_\mathrm{S}\mathcal{T}_\mathrm{SP}\mathcal{R}_\mathrm{P}\,.
\end{equation}
Here, $\mathcal{R}_\mathrm{P}$ and $\mathcal{R}_\mathrm{S}$ denote the reflection operator at the plane and sphere, respectively, while $\mathcal{T}_{\mathrm{PS}}$ stands for the translation operator from the sphere center to the surface of the plane and vice versa for $\mathcal{T}_{\mathrm{SP}}$.

The scattering formula for the Casimir force $F$ can be found from \eqref{eq:calF} by taking the negative derivative with respect to the surface-to-surface separation $z$. Application of Jacobi's formula then yields
\begin{equation}\label{eq:F}
F = -\frac{\partial \mathcal{F}}{\partial z}=k_B T \sum_{l=0}^\infty{}'
{\rm tr}\left(\frac{\partial_z \mathcal{M}}{1-\mathcal{M}}\right)\,.
\end{equation}

The round-trip operator \eqref{eq:round-trip_operator} and its constituents act on electromagnetic modes which are solutions of the Helmholtz equations. Through the determinant in \eqref{eq:calF} and the trace in \eqref{eq:F} it is evident that one is free to choose a basis in which those electromagnetic modes are expanded.
Usually, spherical multipoles are employed~\cite{MaiaNeto2008,Emig2008,Canaguier-Durand2010,Hartmann2017}. Recently, it was shown that a basis consisting of plane waves shows far better convergence rates \cite{28}.
Here, we thus employ the approach based on plane waves. In the following, we review the plane-wave numerical approach from \cite{28} and give the relevant ingredients for the
sphere-plate problem.

\subsection{Plane-Wave Representation}

Within the angular spectrum decomposition, we define the plane-wave basis by the set $\{\ket{\mathbf{k}, \Upsilon, \varrho}\}$ with
the component of the wave vector transverse to the $z$-axis $\mathbf{k}$, the propagation direction with respect to the $z$-axis
$\Upsilon=\pm$ and $\varrho=\mathrm{TE},\mathrm{TM}$ denoting transverse electric and transverse magnetic polarizations, respectively.
Even though the plane-wave basis elements also depend on the imaginary frequency $\xi$, we suppress it as $\xi$ is conserved throughout
a round-trip.

When expanding in the plane-wave basis, each operator $\cal O$ appearing in \eqref{eq:round-trip_operator} becomes an integral operator defined through its corresponding scalar kernel function $f_{\cal O}$. These kernel functions correspond to the plane-wave matrix elements of the respective operators. For instance, the action of the round-trip operator on a plane-wave basis element can be written as
\begin{equation}\label{eq:f_M}
\mathcal{M}\ket{\mathbf{k}, -,  \varrho} = \sum_{ \varrho'}\int\frac{d^2{k}'}{(2\pi)^2}
 f_\mathcal{M}(\mathbf{k}',  \varrho';\mathbf{k},  \varrho) \ket{\mathbf{k}', -,  \varrho'}
\end{equation}
in terms of the kernel function of the round-trip operator $f_\mathcal{M}$.

The translation operators and the reflection operator on the plane are diagonal in the plane-wave basis. Their corresponding kernel functions read
\begin{align}
f_{\mathcal{T}_\mathrm{PS}}(\mathbf{k}',  \varrho';\mathbf{k},  \varrho) &=
 f_{\mathcal{T}_\mathrm{SP}}(\mathbf{k}',  \varrho';\mathbf{k},  \varrho) =
 (2\pi)^2 e^{-\kappa (z+R)} \delta_{ \varrho, \varrho'}\delta(\mathbf{k}-\mathbf{k}'), \nonumber \\
f_{\mathcal{R}_\mathrm{P}}(\mathbf{k}',  \varrho';\mathbf{k},  \varrho) &= (2\pi)^2 r_{ \varrho}
\delta_{ \varrho, \varrho'} \delta(\mathbf{k}-\mathbf{k}')
\end{align}
\noindent
with
\begin{equation}\label{eq:kappa}
\kappa = \sqrt{K^2 + \mathbf{k}^2}\,,
\end{equation}
\noindent
where the imaginary vacuum wave number $K=\xi/c$ and the Fresnel reflection
coefficients are
\begin{equation}\label{Fresnel}
\begin{aligned}
r_\mathrm{TE} &= \frac{\kappa - \sqrt{(\varepsilon-1)K^2 + \kappa^2}}{\kappa + \sqrt{(\varepsilon-1)K^2 + \kappa^2}},\\
r_\mathrm{TM} &= \frac{\varepsilon\kappa - \sqrt{(\varepsilon-1)K^2 + \kappa^2}}{\varepsilon\kappa + \sqrt{(\varepsilon-1)K^2 + \kappa^2}}\,.
\end{aligned}
\end{equation}
\noindent
The Fresnel reflection coefficients (\ref{Fresnel}) depend explicitly on the dielectric
permittivity $\varepsilon=\varepsilon(i\xi)$.

The reflection on the sphere is only diagonal in the polarization basis defined with respect to the scattering plane. In the TE-TM polarization basis taken with respect to the plate surface,
 the reflection operator is, however, not diagonal.
Within the sphere-plate geometry, the kernel function of the round-trip operator
is thus proportional to the kernel function of the reflection operator on the sphere:
\begin{equation}
f_\mathcal{M}(\mathbf{k}',  \varrho';\mathbf{k},  \varrho) = r_{ \varrho}
 e^{-(\kappa+\kappa')(z+R)} f_{\mathcal{R}_\mathrm{S}}(\mathbf{k}',  \varrho';\mathbf{k},  \varrho),
\end{equation}
where \cite{26,27}
\begin{equation}\label{eq:kernel_RS}
\begin{aligned}
f_{\mathcal{R}_\mathrm{S}}(\mathbf{k}', \mathrm{TM};\mathbf{k}, \mathrm{TM}) &= \frac{2\pi}{K \kappa'}(A S_2 + B S_1), \\
f_{\mathcal{R}_\mathrm{S}}(\mathbf{k}', \mathrm{TE};\mathbf{k}, \mathrm{TE}) &= \frac{2\pi}{K \kappa'}(A S_1 + B S_2), \\
f_{\mathcal{R}_\mathrm{S}}(\mathbf{k}', \mathrm{TM};\mathbf{k}, \mathrm{TE}) &= -\frac{2\pi}{K \kappa'}(C S_1 + D S_2), \\
f_{\mathcal{R}_\mathrm{S}}(\mathbf{k}', \mathrm{TE};\mathbf{k}, \mathrm{TM}) &= \frac{2\pi}{K \kappa'}(C S_2 + D S_1),
\end{aligned}
\end{equation}
with $\kappa'$ defined as in \eqref{eq:kappa} and the plane-wave scattering amplitudes \cite{BohrenHuffman}
\begin{equation}\label{eq:S1S2}
\begin{aligned}
S_1 &= \sum_{\ell=1}^\infty \frac{2\ell+1}{\ell(\ell+1)}
\left[ a_\ell(i K R) \pi_\ell(\cos\Theta) + b_\ell(i K R) \tau_\ell(\cos\Theta)\right],
\\
S_2 &= \sum_{\ell=1}^\infty \frac{2\ell+1}{\ell(\ell+1)}
\left[ a_\ell(i K R) \tau_\ell(\cos\Theta) + b_\ell(i K R) \pi_\ell(\cos\Theta)\right]\,.
\end{aligned}
\end{equation}

These plane-wave scattering amplitudes depend on the material properties of the sphere through the Mie coefficients \cite{BohrenHuffman}
\begin{equation}\label{eq:Mie_coefficients}
\begin{aligned}
a_\ell(ix) &= (-1)^{\ell} \frac{\pi}{2} \frac{n^2 s_\ell^{(a)} - s_\ell^{(b)}}{n^2 s_\ell^{(c)} + s_\ell^{(d)}} \,,\\
b_\ell(ix) &= (-1)^{\ell+1} \frac{\pi}{2} \frac{s_\ell^{(b)} - s_\ell^{(a)}}{s_\ell^{(c)} + s_\ell^{(d)}}
\end{aligned}
\end{equation}
for electric and magnetic polarizations, respectively, where
\begin{equation}\label{eq:s_coefficients}
\begin{aligned}
s_\ell^{(a)} &= I_{\ell+\frac{1}{2}}(nx)\left[x I_{\ell-\frac{1}{2}}(x)-\ell I_{\ell+\frac{1}{2}}(x)\right]\,, \\
s_\ell^{(b)} &=I_{\ell+\frac{1}{2}}(x)\left[nx I_{\ell-\frac{1}{2}}(nx)-\ell I_{\ell+\frac{1}{2}}(nx)\right]\,, \\
s_\ell^{(c)} &= I_{\ell+\frac{1}{2}}(nx)\left[x K_{\ell-\frac{1}{2}}(x)+\ell K_{\ell+\frac{1}{2}}(x)\right]\,, \\
s_\ell^{(d)} &= K_{\ell+\frac{1}{2}}(x)\left[nx I_{\ell-\frac{1}{2}}(nx)-\ell I_{\ell+\frac{1}{2}}(nx)\right]
\end{aligned}
\end{equation}
with $I_\ell$ and $K_\ell$ being the modified Bessel functions of first and second kind~\cite{NIST:DLMF} and
 $n=n(i\xi)=\sqrt{\varepsilon(i\xi)}$ the refractive index of the sphere.
The angular functions $\pi_\ell$ and $\tau_\ell$ appearing in (\ref{eq:S1S2}) are defined as \cite{BohrenHuffman}
\begin{equation}
\begin{aligned}
\pi_\ell(z) &= P_\ell'(z), \\
\tau_\ell(z) &= -(1-z^2)P_\ell''(z) + xP_\ell'(z),
\end{aligned}
\end{equation}
where $ P_\ell(z)$ denotes the Legendre polynomial~\cite{NIST:DLMF}.
The angular functions only depend on the scattering angle $\Theta$ which is expressed as
\begin{equation}\label{eq:scattering_angle}
\cos\Theta = -\frac{\mathbf{k}\cdot \mathbf{k}' + \kappa\kappa'}{K^2}\,.
\end{equation}

The functions $A$, $B$, $C$ and $D$ in (\ref{eq:kernel_RS})
describe a rotation from the polarization basis defined through the scattering plane to the TE-TM polarization basis.
They are functions of the incident and scattered wave vectors and can be expressed as \cite{26}
\begin{equation}\label{eq:ABCD}
\begin{aligned}
A &= \frac{K^4 \cos(\Delta\varphi) - [kk'\cos(\Delta\varphi) -\kappa \kappa'][kk'-\kappa \kappa'\cos(\Delta\varphi)]}{K^4 - [kk'\cos(\Delta\varphi) -\kappa \kappa']^2}\,, \\
B &= -\frac{K^2kk'\sin^2(\Delta\varphi)}{K^4 - [kk'\cos(\Delta\varphi) -\kappa \kappa']^2}\,, \\
C &= K\sin(\Delta\varphi) \frac{kk' \kappa \cos(\Delta\varphi)+k^2 \kappa'}{K^4 - [kk'\cos(\Delta\varphi) -\kappa \kappa']^2}\,, \\
D &= -K\sin(\Delta\varphi) \frac{kk' \kappa' \cos(\Delta\varphi)+{k'}^2 \kappa}{K^4 - [kk'\cos(\Delta\varphi) -\kappa \kappa']^2}\,,
\end{aligned}
\end{equation}
where we have employed polar coordinates $\mathbf{k}=(k,\varphi)$ and $\mathbf{k}'=(k',\varphi')$, and $\Delta\varphi = \varphi'-\varphi$. Note that the fact that the sphere center is located along the positive $z$-direction above the plane is encoded in the sign of the coefficients $C$ and $D$. An opposite orientation of the $z$-axis would flip their signs.

\subsection{Zero-Frequency Limit}

The scattering formulas \eqref{eq:calF} and \eqref{eq:F} require an evaluation of the matrix elements at $\xi_0 = 0$. The zero-frequency limit $\xi\rightarrow 0$ or equivalently $K\rightarrow 0$ of the reflection operators depends on the modeling of the dielectric functions. In the following, we will specifically consider the Drude and the plasma models which are given by
\begin{equation}
\varepsilon_D=\varepsilon_D(i\xi_l)=1+\frac{\omega_p^2}{\xi_l(\xi_l+\gamma)}, \quad
\varepsilon_p=\varepsilon_p(i\xi_l)=1+\frac{\omega_p^2}{\xi_l^2},
\label{epsD}
\end{equation}
\noindent
where $\omega_p$ is the plasma frequency and $\gamma$ is the relaxation parameter.

The zero-frequency limit of the Fresnel reflection coefficients is straightforward. For the Drude model, we find from (\ref{Fresnel})
\begin{equation}
r^{{D}}_\mathrm{TM} = 1\,,\qquad
r^{{D}}_\mathrm{TE} = 0
\end{equation}
while for the plasma model, we obtain
\begin{equation}
r^{{p}}_\mathrm{TM} = 1\,,\qquad
r^{{p}}_\mathrm{TE} =
\frac{\vert \mathbf{k} \vert - \sqrt{K_p^2 + \mathbf{k}^2}}{\vert \mathbf{k} \vert + \sqrt{K_p^2 + \mathbf{k}^2}}
\end{equation}
with the plasma wave number $K_p=\omega_p/c$.

The zero-frequency limit for the kernel functions of the reflection operator on the sphere requires more work. It is easy to see that the polarization transformation coefficients \eqref{eq:ABCD} become
\begin{equation}\label{eq:ABCD_lowfreq}
A = 1\,,\quad B=C=D=0\,,
\end{equation}
and we are left with the task of finding the zero-frequency limit of the scattering amplitudes \eqref{eq:S1S2}.
Note that while the scattering amplitudes
\eqref{eq:S1S2} vanish in the limit $K\rightarrow 0$, this is not the case for the kernel functions \eqref{eq:kernel_RS}.
Therefore, we need to keep terms linear in $K$ in the low-frequency expression for the scattering amplitudes.

We start by expanding the angular functions $\pi_\ell$ and $\tau_\ell$. According to \eqref{eq:scattering_angle}, $\cos\Theta$ diverges like $1/K^2$ at low frequencies. Thus, we can employ the asymptotical expressions of Legendre functions for large arguments \cite{NIST:DLMF}
(see \S14.8) to find
\begin{equation}\label{eq:pitau_lowfreq}
\begin{aligned}
\pi_\ell\big(\cos\Theta\big) &\sim \frac{(2\ell)!}{2^\ell (\ell-1)! \ell!} \cos^{\ell-1}\Theta
	\propto\frac{1}{K^{2\ell-2}}\,, \\
\tau_\ell\big(\cos\Theta\big) &\sim \frac{(2\ell)!}{2^\ell [(\ell-1)!]^2} \cos^{\ell}\Theta
	\propto\frac{1}{K^{2\ell}}\,.
\end{aligned}
\end{equation}
As a consequence, among the four combinations of these two functions and the two Mie coefficients $a_\ell$ and $b_\ell$, only those involving $\tau_\ell$ can potentially lead to contributions linear in $K$. Terms involving $\pi_\ell$ yield an additional factor $K^2$ and can thus be disregarded.

At low frequencies, the Mie coefficient are of the form (see \S7 of \cite{Canaguier2011}
for a detailed discussion)
\begin{equation}\label{eq:mie_lowfreq}
\begin{aligned}
a_\ell(iKR) &= (-1)^{\ell} \frac{(\ell+1)(\ell!)^2}{2\ell(2\ell+1)[(2\ell)!]^2} (2KR)^{2\ell+1} + \mathcal{O}\left((KR)^{2\ell+2}\right)\,, \\
b_\ell(iKR) &= (-1)^{\ell} \frac{(\ell+1)(\ell!)^2}{2\ell(2\ell+1)[(2\ell)!]^2} \mathcal{B}_\ell^\mathrm{model} (2KR)^{2\ell+1} + \mathcal{O}\left((KR)^{2\ell+2}\right)\,,
\end{aligned}
\end{equation}
where the coefficient $\mathcal{B}_\ell^\mathrm{model}$ depends on the model used for the material under consideration.
For the Drude model, we have
\begin{equation}
\mathcal{B}_\ell^{D} = 0
\end{equation}
and, for the plasma model, we have
\begin{equation}
\mathcal{B}_\ell^{p} = -\frac{\ell}{\ell+1}\left[1- \frac{2\ell+1}{K_p R} \frac{I_{\ell+1/2}(K_p R)}{I_{\ell-1/2}(K_p R)}\right].
\end{equation}

With the low-frequency asymptotic expressions \eqref{eq:pitau_lowfreq} and \eqref{eq:mie_lowfreq}, the scattering amplitudes in the low-frequency limit read
\begin{equation}\label{eq:S1S2_lowfreq}
\begin{aligned}
S_1 & =  KR \sum_{\ell=1}^\infty \mathcal{B}_\ell^\mathrm{model} \frac{[2 R^2(\mathbf{k}\cdot \mathbf{k}' + \vert \mathbf{k}\vert \vert \mathbf{k}'\vert)]^\ell}{(2\ell)!}\,, \\
S_2 & =  KR \sum_{\ell=1}^\infty  \frac{[2 R^2(\mathbf{k}\cdot \mathbf{k}' + \vert \mathbf{k}\vert \vert \mathbf{k}'\vert)]^\ell}{(2\ell)!}\,.
\end{aligned}
\end{equation}

Inserting \eqref{eq:ABCD_lowfreq} and \eqref{eq:S1S2_lowfreq} in \eqref{eq:kernel_RS}, the low-frequency limit of the kernel functions can finally be expressed as
\begin{equation}
\begin{aligned}
f_{\mathcal{R}_\mathrm{S}}(\mathbf{k}', \mathrm{TM};\mathbf{k}, \mathrm{TM}) &= \frac{2\pi R}{k'} \sum_{\ell=1}^\infty  \frac{[2 R^2(\mathbf{k}\cdot \mathbf{k}' + \vert \mathbf{k}\vert \vert \mathbf{k}'\vert)]^\ell}{(2\ell)!},\\
f_{\mathcal{R}_\mathrm{S}}(\mathbf{k}', \mathrm{TE};\mathbf{k}, \mathrm{TE}) &= \frac{2\pi R}{k'} \sum_{\ell=1}^\infty \mathcal{B}_\ell^\mathrm{model} \frac{[2 R^2(\mathbf{k}\cdot \mathbf{k}' + \vert \mathbf{k}\vert \vert \mathbf{k}'\vert)]^\ell}{(2\ell)!}, \\
f_{\mathcal{R}_\mathrm{S}}(\mathbf{k}', \mathrm{TM};\mathbf{k}, \mathrm{TE}) &= 0 ,\qquad
f_{\mathcal{R}_\mathrm{S}}(\mathbf{k}', \mathrm{TE};\mathbf{k}, \mathrm{TM}) = 0\,.
\end{aligned}
\end{equation}
The kernel function  $f_{\mathcal{R}_\mathrm{S}}(\mathbf{k}', \mathrm{TM};\mathbf{k}, \mathrm{TM})$ is the same for both
 models. Moreover, the sum over $\ell$ can be performed in this case~\cite{Schoger2020}, leading to
\begin{equation}
f_{\mathcal{R}_\mathrm{S}}(\mathbf{k}', \mathrm{TM};\mathbf{k}, \mathrm{TM}) = \frac{2\pi R}{k'}\left[\cosh\left(R\sqrt{2(\mathbf{k}\cdot \mathbf{k}' + \vert \mathbf{k}\vert \vert \mathbf{k}'\vert)}\right)-1\right]\,.
\end{equation}

\subsection{Numerical Application}

To make the evaluation of the scattering formula within the plane-wave basis amenable to numerical calculations, a discretization of the continuous wave vectors is required. In order to exploit the cylindrical symmetry of the problem, we employ polar coordinates $\mathbf{k}=(k, \varphi)$. The two integrals over radial and angular components of the in-plane wave vector in \eqref{eq:f_M} can then be discretized using one-dimensional quadrature rules. Denoting the quadrature nodes and weights for the radial and angular components as $(k_i, w_i)$ for $i=1,\ldots, N$ and $(\varphi_j, v_j)$ for $j=1,\ldots,M$, respectively, we can express \eqref{eq:f_M} in discretized form as
\begin{equation}\label{eq:f_M_discrete}
\mathcal{M}\ket{k_i, \varphi_j, -, \varrho} = \sum_{\varrho'=\mathrm{TE},\mathrm{TM}}
\sum_{i'=1}^N\sum_{j'=1}^M \frac{ k_{i'} w_{i'} v_{j'}}{(2\pi)^2}
 f_\mathcal{M}(k_{i'}, \varphi_{j'}, \varrho';k_{i}, \varphi_j, \varrho) \ket{k_{i'}, \varphi_{j'}, -, \varrho'}
\end{equation}
for $\varrho=\mathrm{TE},\mathrm{TM}$. Consequently, the discretized matrix elements of the round-trip operator read \cite{Bornemann2010}
\begin{equation}\label{eq:roundtrip_op_discrete}
\braket{k_{i'}, \varphi_{j'}, -, \varrho' \vert\mathcal{M}\vert k_i, \varphi_j, -, \varrho} =
\frac{ k_{i'} w_{i'} v_{j'}}{(2\pi)^2} f_\mathcal{M}(k_{i'}, \varphi_{j'}, \varrho';
k_{i}, \varphi_j, \varrho)\,.
\end{equation}
The finite matrix \eqref{eq:roundtrip_op_discrete} is a threefold block matrix with
respect to the indices of radial and angular quadrature rule and the polarization component.

The scattering formulas for the Casimir free energy \eqref{eq:calF} and Casimir force \eqref{eq:F} can now be evaluated by replacing the round-trip operator with the corresponding finite matrix \eqref{eq:roundtrip_op_discrete}. While in principle one is free to choose quadrature rules, we found those specified in the following particularly suited for the problem at hand \cite{28}.

As the quadrature rule for the radial component we employ the Fourier-Chebyshev quadrature rule introduced in \cite{Boyd1987}. With
\begin{equation}
t_i = \frac{\pi i}{N+1}\,,
\end{equation}
the quadrature rule is specified by its nodes
\begin{equation}\label{eq:quadrature-points}
k_i = a \cot^2\frac{t_i}{2}
\end{equation}
and weights
\begin{equation}
w_i = \frac{8 a \sin t_i}{[1-\cos t_i]^2}\frac{1}{N+1} \sum_{\substack{ j=1 \\\text{$j$ odd}}}^N \frac{\sin(j t_i)}{j}
\end{equation}
for $i=1,\dots,N$. An optimal choice for the
free parameter $a$ can boost the convergence of the computation.
For dimensional reasons, the transverse wave vector and thus $a$ should scale
like the inverse surface-to-surface distance $1/z$. In fact, the choice $a=1/z$
already yields a fast convergence rate as was demonstrated in \cite{28}.

For the angular component of the in-plane wave vectors we employ the trapezoidal quadrature
rule. Its nodes and weights are defined by $\varphi_j = 2\pi j/M$ and $v_j=2\pi/M$,
respectively, for $j=1,\ldots,M$. While for arbitrary functions the trapezoidal rule
is not efficient, it is exponentially convergent for periodic functions appearing here.

Moreover, the trapezoidal rule allows us to further exploit the cylindrical symmetry of
the problem. Note that due to the cylindrical symmetry the kernel function of the
round-trip operator \eqref{eq:f_M} depends only on the difference $\Delta\varphi = \varphi'-\varphi$
 of angular components. Using the trapezoidal rule, the discretized
matrix elements \eqref{eq:roundtrip_op_discrete} then only depend on the difference of
indices $j'-j$. A matrix of this form is known as circulant block matrix. It is well-known
that a circulant block matrix can be block diagonalized using a discrete Fourier transform,
thus reducing the complexity of the problem.

In fact, the indices labeling the diagonal blocks after the discrete Fourier transform can
then be identified with the angular indices $m,$ known from the spherical multipole
representation, which denote the axial component of the electromagnetic field angular
momentum. Particularly at short distances, the plane-wave basis is advantageous with respect
to the spherical multipole basis because the required size of the block matrices is
considerably smaller as the following estimate demonstrates. The size of the matrices is
determined by the radial quadrature order $N$ in the plane-wave representation. It can be
shown that in order to maintain a certain numerical error, it needs to scale as
$N\propto \sqrt{R/z}$ \cite{28}. On the other hand, within the spherical
multipole representation the block matrix size is determined by the highest multipole
index $\ell_\mathrm{max}$ included in the calculation which scales linearly in $R/z$.

The obtained computational results are shown below in Section 6.2 (Figures \ref{figT4} and
\ref{figT5}).
Note that at all nonzero Matsubara frequencies the dielectric permittivities
$\varepsilon(i\xi_l)$
determined from the optical data of Au extrapolated to zero frequency by means of either the
Drude or the plasma model \cite{ourBook} have been used in computations.


\section{Computation of the Casimir Force in Sphere-Plate Geometry Based on the
Derivative Expansion}

\newcommand{\be}{\begin{equation}}
\newcommand{\ee}{\end{equation}}
 \newcommand{\kk}{{k_0}^2-k^2}
 \definecolor{BrickRed}{cmyk}{0,0.89,0.94,0.28}
\definecolor{MidnightBlue}{cmyk}{0.98,0.13,0,0.43}
\definecolor{DarkGreen}{rgb}{0,0.7,0.1}
\newcommand{\add}[1]{{\color{magenta} #1}}
\newcommand{\com}[1]{{\color{MidnightBlue}\{\small \sc #1\}}}
\newcommand{\del}[1]{{\color{red}[[#1]]}}

The   Casimir force between a sphere and a plate has been also computed using a different
approach, based on the derivative expansion (DE) \cite{29,30,31,fosco2,fosco3,bimonteprecise,bimonte2sphere}.
The starting point is the following  decomposition of the Casimir force:
\be
{ F}={ F}_{l=0}+{ F}_{l>0}\;,\label{forsplit}
\ee
where ${ F}_{l=0}$ represents the contribution of the $l=0$ Matsubara frequency (the so-called classical contribution), while ${ F}_{l>0}$ represents the contribution of the non-vanishing Matsubara frequencies $\xi_l$ with $l$=1,\,2,\,\dots
Within the Drude model ${ F}_{l=0}$ receives a contribution only from transverse magnetic (TM) polarization:
\be
{ F}_{l=0} \vert_{\rm Drude}={ F}_{l=0}^{(\rm TM)}\;.
\ee
The Drude-model classical force ${ F}_{l=0}^{(\rm TM)}$ for the sphere-plate geometry has been computed  analytically in  \cite{bimonteex1}.
Within the plasma model, the classical term receives a contribution also from transverse electric (TE) polarization:
\be
{ F}_{l=0}\vert_{\rm plasma}={ F}_{l=0}^{(\rm TM)}+{ F}_{l=0}^{(\rm TE)}\vert_{\rm plasma}\;,
\ee
where the plasma-model TM contribution coincides with the corresponding term of the Drude model. The TE classical term ${ F}_{l=0}^{(\rm TE)}\vert_{\rm plasma}$ cannot be computed analytically,  but it can be  computed  numerically with high precision by using the scattering formula
discussed in the previous section.

For both the Drude model and the plasma model, the contribution ${ F}_{l>0}$ of the
non-vanishing Matsubara modes can be computed very accurately by using a semi-analytical
approach based on the DE, as we now explain.
Instead of a sphere in front of a plate, consider a more general gently curved dielectric
surface, described by a smooth height profile $z=H(x,y)$, where $(x,y)$ are cartesian
coordinates spanning the plate surface $\Sigma$, and the $z$ axis is drawn perpendicular
to the plate towards the surface.   The starting point of the  DE is the assumption that
the Casimir force $F_{l>0}$ admits a  {\it local} expansion in {\it powers of derivatives}
of the  height profile $H$:
\be
{ F}_{l>0}={ F}_{l>0}^{(\rm PFA)} + \int_{\Sigma}  d^2 x \; X(H)  (\nabla H)^2 + \rho^{(2)}\;,\label{derexp}
\ee
where ${F}_{l>0}^{(\rm PFA)}$ is the familiar PFA expression (restricted  to the
 contributions with $l>0$), and  $X(H)$ is a function to be determined.

The quantity  $\rho^{(2)}$ represents corrections that become negligible as the local
 radius of curvature of the surface $R$ goes to infinity  for fixed minimum surface-plate
 distance $z$. The function $X(H)$  is determined by matching the DE with the
 perturbative expansion   of the Casimir force, in the common domain of validity of the
 two expansions. The matching procedure leads to an expression for $X$, having the
 form of an integral over the in-plane momentum, that can be easily computed numerically
 (for details, see \cite{29,30,31}). The validity of the ansatz made
 in (\ref{derexp})  hinges on the {\it locality} properties of the Casimir force.
The key point   is that for imaginary frequencies $\omega={\rm i} \xi_l$ with $l>0$,
 photons acquire an effective mass proportional to $l$, which renders the interaction more and more local as $l$ increases.
 This argument leads one to expect that (\ref{derexp}) becomes more and more accurate,
 as $l$ increases.  We recall that, prior to the discovery in  \cite{bimonteex1} of the exact
 expression of the Drude-model classical term,  the DE was used in \cite{29,30} to
 compute  curvature corrections to the Drude-model Casimir force between Au sphere and plate
 at room temperature. In this work, the DE is restricted to the massive $l>0$ terms.
 When the integral in (\ref{derexp}) is evaluated for a sphere with
$H(x,y)=z+(x^2+y^2)/(2 R)+ \dots$,  and only terms of  order $z/R$ are retained, one ends up with an expression that can be recast in the form:
\be
{ F}_{l>0}={ F}_{l>0}^{(\rm PFA)} \left(1- \theta \; \frac{z}{R} \right)\;.\label{DEF}
\ee
where the dimensionless coefficient $\theta$  can be expressed in terms of  the parameter
 $X$ in \\(\ref{derexp}) (see \cite{bimonte2sphere}).
 The coefficient $\theta$   depends on the separation $z$, on the temperature $T$ and of course  on
the permittivities $\varepsilon_l=\varepsilon({\rm i} \xi_l)$, but it is independent of the sphere radius $R$.
The Drude and plasma-model values of  $\theta$  for gold can be found in \cite{bimonte2sphere}.
 Combining (\ref{derexp}) with (\ref{DEF}),  we  obtain the following   expression for $F$:
\be
{ F}_{\rm DE} ={ F}_{l=0}+{ F}_{l>0}^{(\rm PFA)} \left(1- \theta \; \frac{z}{R} \right)\;.\label{forDE}
\ee

The reader may wonder at this point why  the DE was not used to estimate as well the  TE
classical term ${ F}_{l=0}^{(\rm TE)}\vert_{\rm plasma}$ of the plasma model. The reason
is that the DE is not valid for this term, due to its nonlocal features. Detailed numerical
and analytical investigations \cite{antoine, bimontehT}  in the limit of perfectly
conducting  sphere and plate, indeed show that the leading correction  beyond PFA in the
small distance expansion of ${ F}_{l=0}^{(\rm TE)}\vert_{\rm plasma}$  is a logarithmic
term $\sim \log(z/R)$, which cannot be described by the DE. As a result, the term
${ F}_{l=0}^{(\rm TE)}\vert_{\rm plasma}$ still needs to be computed numerically using
the scattering formula.

\begin{figure}[!h]
\hspace*{1cm}
\includegraphics[width=.7\columnwidth]{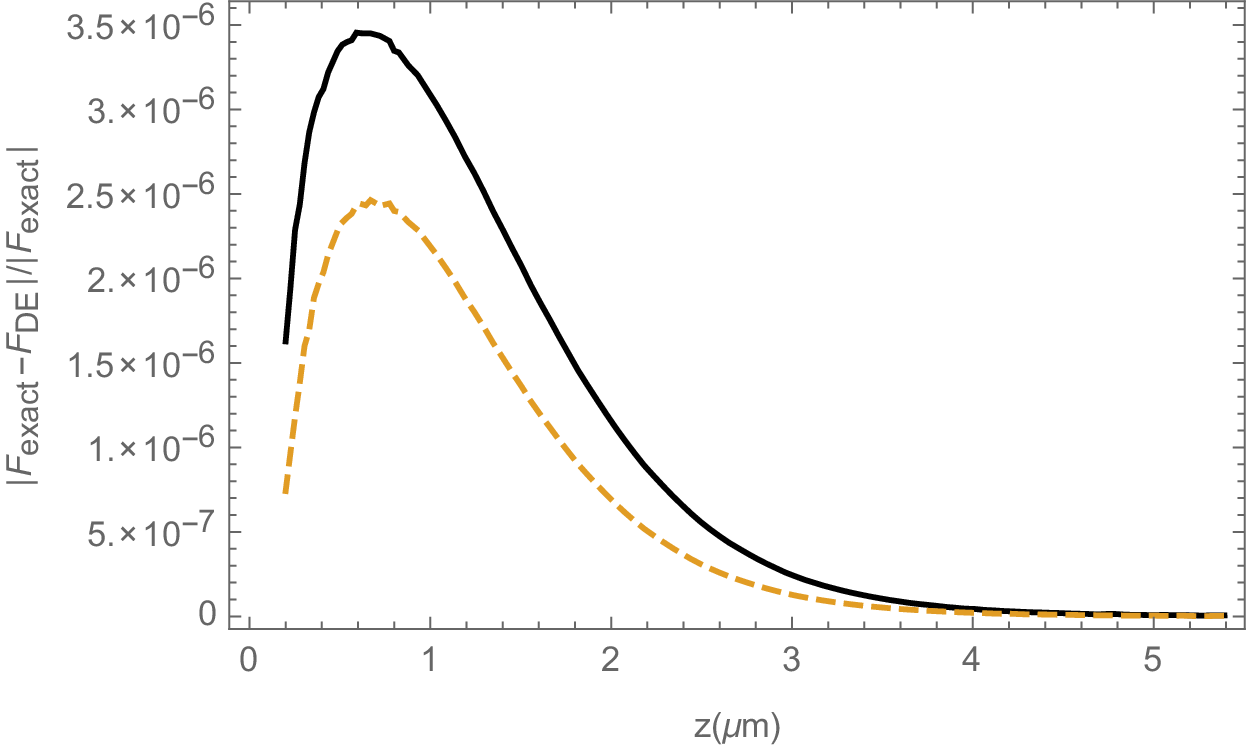}
\caption{\label{errorDE}  Fractional difference $\eta=|F_{\rm exact}-F_{\rm DE}|/|F_{\rm exact}|$ between the sphere-plate exact Casimir force $F_{\rm exact}$ and  the approximate formula (\ref{forDE}), obtained from the DE for the nonzero Matsubara frequencies and exact result for
the zero-frequency contribution.
Sphere radius $R=149.7~\mu$m and  $T=295.25$~K.
The solid and dashed lines  are for the Drude and the plasma models, respectively.}
\end{figure}

In order to assess quantitatively the precision of the approximate formula (\ref{forDE}),
we have compared the values of the Casimir force provided by (\ref{forDE})  with those obtained
 by the high-precision numerical computation of the scattering formula in Section~4.
The agreement between the respective values of the force is excellent,  as it is demonstrated
by Figure~\ref{errorDE} which shows a plot of  the fractional difference
$\eta=|F_{\rm exact}-F_{\rm DE}|/|F_{\rm exact}|$
between the  two estimates of the Casimir force, in the separation range extending from 200~nm to $5.4~\mu$m, for  the
sphere radius $R=149.7~\mu$m and for $T=295.25$~K. The solid and dashed lines in Figure~\ref{errorDE} are for the Drude and
the plasma models, respectively. The plot shows that for separations smaller than approximately $0.6~\mu$m,  $\eta$  decreases as
the separation decreases. This is consistent with the fact that the DE is exact in the limit of vanishing separation. The decrease displayed
by $\eta$ for larger values of the separation $z$   is explained by the fact that as the separation increases, the Casimir force is more
and more dominated by the classical term, and therefore the error in the contribution of the non-vanishing  Matsubara frequencies
becomes less and less important. The plot shows that for all displayed separations  $\eta < 3.5 \times 10^{-6}$ within the Drude model,
 while for the plasma
model    $\eta < 2.5 \times 10^{-6}$.  The remarkable agreement between the values of the Casimir force provided by (\ref{forDE})  with those obtained  by direct numerical evaluation of the scattering formula   demonstrates the high  precision of the theoretical predictions of the Casimir force provided by the two methods of computation.

\section{Total Errors and the Comparison between Experiment and Theory}

In this section, we determine the random errors, total systematic errors, and consider two methods
for presentation of the measurement data with estimated role of patch potentials. Then the
experimental results are compared with the theoretical Casimir forces computed in Sections~4
and 5 using the extrapolation of the optical data by means of the Drude and plasma models.

\subsection{Random and Total Experimental Errors}

There are different methods for estimating  the true value of measured physical
quantity and of the respective confidence interval at the desired confidence
probability. In the  simplest cases the individual values of some random quantity obey
either the normal or the homogeneous distribution law. If, however, the factual
distribution law deviates from both the normal and the homogeneous ones, the
estimation of the true values using these laws may lead to unrealistic results.

\begin{figure}[!b]
\centering
\includegraphics[width=.8\columnwidth]{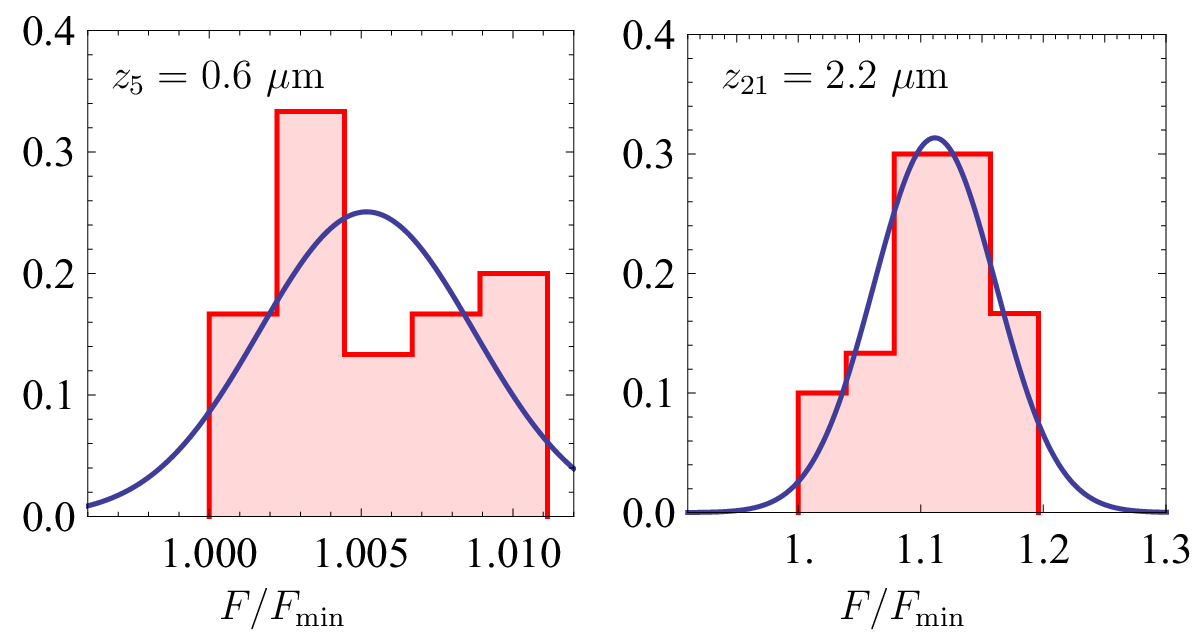}
\caption{The histograms at $z_{5}=0.6~\mu$m (left) and
$z_{21}=2.2~\mu$m (right).
\label{figT1}}
\end{figure}

In our case, the Casimir force was independently measured for thirty times at
the separation distances $z_k$ where $1\leqslant k\leqslant 79$. This means that
at each $z_k$ one should check the character of the distribution law for $n=30$
force values before estimating the true value of the force at each specific
separation.
Close inspection of the measurement data shows that at some separations $z_k$
their distribution law deviates from the normal one.
As an example,
in Figure~\ref{figT1} (left) we present the histogram
at the separation $z_5=0.6~\mu$m. On the $x$-axis we plot the force divided
by the minimum force value at this separation. On the $y$-axis we plot the
normalized value of the probability. It is clearly seen that the distribution
law in this case is far from being normal. In so doing the exact
form of the distribution remains unknown.

There are also separation distances where the distribution law is very close to the
normal one. The relevant histogram at the separation  $z_{21}=2.2~\mu$m is shown
in Figure~\ref{figT1} (right). Using the $\chi^2$  method of verification of hypotheses,
we have found that the hypothesis of a normal distribution does not contradict to our
measurement data at a 75\% confidence level. This, however, does not allow to use the
normal distribution in the analysis of random errors and employ the mean value of the
force $\bar{F}^{\rm expt}(z_k)$ as an estimation of its true value. If the hypothesis
of a normal distribution were proven, one could make a solid statement that at the
confidence probability $\beta$ the measured quantity belongs to the confidence interval
$[\bar{F}^{\rm expt}-\Delta_{\beta},\bar{F}^{\rm expt}+\Delta_{\beta}]$ where
$\Delta_\beta$ is determined by the normal law.
In our case, however, the normal distribution is not a solid fact, but only a hypothesis
which does not contradict to the data at the confidence level of 75\%.
Then the above statement would be unjustified.

Fortunately, mathematical
statistics elaborated special (median) method on how to deal with this problem
\cite{T1}. The median method allows finding an estimation of the  true value of
the measured quantity even if the distribution law is unknown. This estimation
is approximately equal to the standard mean value if the distribution is close
to the normal one. Otherwise, the median method provides an alternative
estimation of the true value which is robust relative to deviations from the
normal law and to the presence of outlying results \cite{T2}.
This method also provides the confidence interval for the estimation of the
true value at the confidence probability $\beta$ (below we consider $\beta=0.95$).
If the distribution is close to  normal, half of the confidence interval
length is approximately equal to the error of the mean found by using the
normal distribution. As applied to our data sets, the median method reduces to
the following \cite{T1}.

Let us arrange the forces measured at the fixed separation $z_k$
 in the order of increasing values
\begin{equation}
F_1^{\rm expt}(z_k)\leqslant F_2^{\rm expt}(z_k)\leqslant\ldots
\leqslant F_{n=30}^{\rm expt}(z_k).
\label{eqT1}
\end{equation}
\noindent
Then, according to the median method, an estimation for the true value of the
force magnitude at the separation $z_k$ is given by the expression
\begin{equation}
\tilde{F}^{\rm expt}(z_k)=\frac{1}{2}\left[F_{\frac{n}{2}=15}^{\rm expt}(z_k)
+F_{\frac{n}{2}+1=16}^{\rm expt}(z_k)\right],
\label{eqT2}
\end{equation}
\noindent
where we take into account that $n=30$ is an even number. Recall also that most of
force values are negative which corresponds to an attraction (with exception of separation
distances exceeding 5.9~$\mu$m where some of the measued forces have positive values,
see the inset in Figure 3).

The confidence interval for an estimation of the true value (\ref{eqT2})
at the chosen confidence probability
$\beta$ is given by
\begin{equation}
\Big(F_{i=i(n,\beta)}^{\rm expt}(z_k),\,
F_{j=j(n,\beta)}^{\,\rm expt}(z_k)\Big),
\label{eqT3}
\end{equation}
\noindent
where the forces $F_{i}^{\,\rm expt}(z_k)$ and
$F_{j}^{\,\rm expt}(z_k)$ belong to the sequence (\ref{eqT1}).
The specific expression for $i$ is given by
\begin{equation}
i=i(n,\beta)={\rm int}\frac{n+1-t_{\beta}\sqrt{n}}{2},
\label{eqT4}
\end{equation}
\noindent
where the symbol int stands for the integer part of the number and $t_{\beta}$ is a
tabulated coefficient. In a similar way,
\begin{equation}
j=j(n,\beta)=1+{\rm int}\frac{n+1+t_{\beta}\sqrt{n}}{2}.
\label{eqT5}
\end{equation}

\begin{figure}[!t]
\centering
\includegraphics[width=.8\columnwidth]{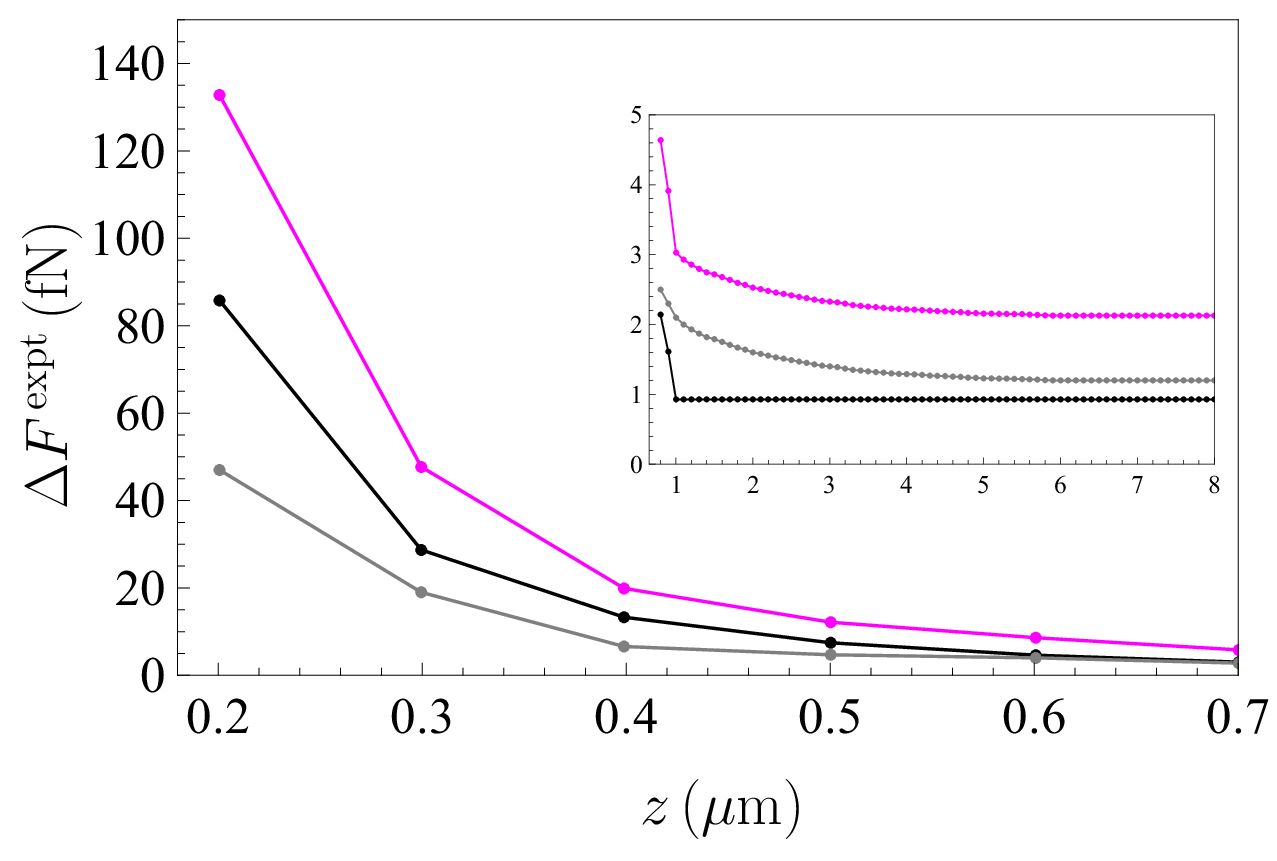}
\caption{Random , systematic  and total  experimental errors in the
estimation of the true values $\tilde{F}^{\rm expt}(z_k)$ are shown by
the gray, black, and pink dots, respectively, as functions of separation.
The separation interval from 0.7 to $8~\mu$m, where all errors vary relatively
slowly, is shown in the inset.
\label{figT2}}
\end{figure}

We calculate all errors at the 95\% confidence level ($\beta=0.95$)
resulting in $t_{0.95}=1.96$ \cite{T1}. Then from (\ref{eqT4}) and (\ref{eqT5})
one obtains $i=10$ and $j=21$. In the framework of the median method, the random
error $\Delta^{\!\rm rand}\tilde{F}^{\rm expt}(z_k)$ is defined as one half of the length of
confidence interval (\ref{eqT3}) with the corresponding smoothing procedure \cite{T1}.
It is shown by the gray line in Figure~\ref{figT2}.

It is interesting to compare the estimation of the true force values using a hypothesis
of the normal distribution and the median method in different cases. We begin with the
separation $z_5=0.6~\mu$m where the distribution of Figure~\ref{figT1} (left) is far
from being normal.  If one, nevertheless, assumes that
it is normal, the following estimation for the true force value, the
confidence interval and the random error are obtained:
\begin{equation}
\bar{F}^{\rm expt}(z_5)= -1662.86\,\mbox{fN},\quad     (-1665.0\,\mbox{fN},-1660.7\,\mbox{fN}),
\quad   \Delta^{\!\rm rand}\bar{F}^{\rm expt}(z_5)=2.1\,\mbox{fN}.
\label{eqT6a}
\end{equation}
\noindent
If the median method is used in this case, which makes the proper account of the
deviations from the normal law, one finds
\begin{equation}
\tilde{F}^{\rm expt}(z_5)= -1666.5\,\mbox{fN},\quad     (-1666.5\,\mbox{fN}, -1658.85\,\mbox{fN}),
\quad   \Delta^{\!\rm rand}\tilde{F}^{\rm expt}(z_5)=3.8\,\mbox{fN}.
\label{eqT6b}
\end{equation}
\noindent
It is seen that the normal distribution underestimates both the true value of the force
and the random error.

Now we consider the separation $z_{21}=2.2~\mu$m where the distribution is rather close
to the normal [see Figure~\ref{figT1} (right)].
The normal distribution gives the following estimation
for the true force value, the confidence interval and the random error:
\begin{equation}
\bar{F}^{\rm expt}(z_{21})= -51.69\,\mbox{fN},\quad
(-52.52\,\mbox{fN},-50.84\,\mbox{fN}),
\quad   \Delta^{\!\rm rand}\bar{F}^{\rm expt}(z_{21})=0.85\,\mbox{fN}.
\label{eqT6c}
\end{equation}
\noindent
The median method for the same data point results in
\begin{equation}
\tilde{F}^{\rm expt}(z_{21})=- 51.67\,\mbox{fN},\quad
(-52.80\,\mbox{fN},-50.97\,\mbox{fN}),
\quad   \Delta^{\!\rm rand}\tilde{F}^{\rm expt}(z_{21})=0.92\,\mbox{fN}.
\label{eqT6d}
\end{equation}
It is seen that differences between the results obtained using both
methods are insignificant.

\begin{figure}[!t]
\centering
\includegraphics[width=.8\columnwidth]{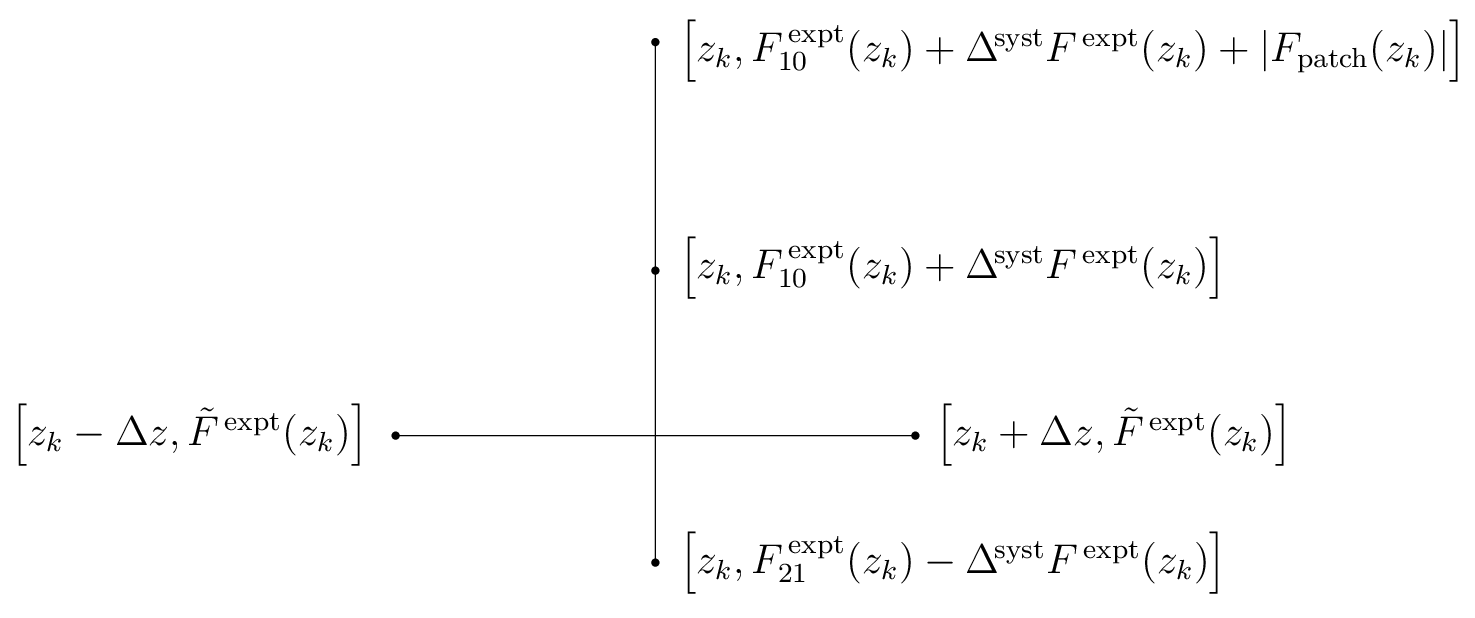}
\caption{The experimental cross at separation $z_k$ taking into consideration the
confidence interval, systematic errors (both determined at the 95\% confidence level),
and the role of patch potentials.
\label{figT3}}
\end{figure}
Below, for
illustration of the estimation (\ref{eqT2})  of the true force values and their
experimental errors and uncertainties including the role of surface patches,
we use crosses centered at the points
$\left[z_k,\tilde{F}^{\,\rm expt}(z_k)\right]$. The meaning of these crosses is
illustrated in Figure \ref{figT3}.
Thus, the upper and lower vertical  arms finish at the points
\begin{eqnarray}
&&
[z_k,F_{10}^{\,\rm expt}(z_k)+\Delta^{\!\rm syst}F^{\,\rm expt}(z_k)
+|F_{\rm patch}(z_k)|],
\nonumber \\
&&
[z_k,F_{21}^{\,\rm expt}(z_k)-\Delta^{\!\rm syst}F^{\,\rm expt}(z_k)],
\label{T11}
\end{eqnarray}
\noindent
where $\Delta^{\!\rm syst}F^{\,\rm expt}(z_k)$ is the total systematic error in
measuring the Casimir force at the separation $z_k$ and $|F_{\rm patch}(z_k)|$ is
an estimated magnitude of the force due to the patch potentials taken from
Figure~\ref{figPatch}. It corresponds to an attraction and, thus, makes an impact on only
the upper vertical arms.

The total systematic error is a combination
of the calibration error $\Delta_{\rm cal}=0.2~$fN, detection error
$\Delta_{\rm det}=0.6~$fN, and separation-dependent error of the
measurement method $\Delta_{\rm meas}$ which includes the role of
thermal/vibration noise. The combination law is given by \cite{ourBook,T1,T4}
\begin{equation}
\Delta^{\!\rm syst}F^{\,\rm expt}(z_k)=\min\left(\Delta_{\rm cal}+\Delta_{\rm det}+
\Delta_{\rm meas},\,k_{\beta}(N)
\sqrt{\Delta_{\rm cal}^2+\Delta_{\rm det}^2+\Delta_{\rm meas}^2}\right),
\label{eqT6}
\end{equation}
\noindent
where for $N=3$ errors
the tabulated coefficient $k_{0.95}(3)=1.11$. Note that the dominant contribution
to (\ref{eqT6}) is given by $\Delta_{\rm meas}$. The quantity
$\Delta^{\!\rm syst}F^{\,\rm expt}(z_k)$ is shown by the black line in
Figure~\ref{figT2}.

Then the total experimental error is given by
\begin{equation}
\Delta\tilde{F}^{\,\rm expt}(z_k)=\Delta^{\!\rm rand}\tilde{F}^{\,\rm expt}(z_k)
+\Delta^{\!\rm syst}{F}^{\,\rm expt}(z_k).
\label{eqT7a}
\end{equation}
\noindent
It is shown by the pink line in Figure~\ref{figT2}.

The horizontal arms of the crosses are equal to the systematic error in measuring the
separation distances since the random one arising due to an averaging over
$n=30$ measurements turns out to be negligibly small as compared to the systematic.
The latter is a combination of the error in $D_1+D_2$,
$\Delta_D=0.6~$nm, the error associated with a flatness of the wafer
$\Delta_{\rm flat}=1.2~$nm and $\Delta_{\rm meas}=0.2~$nm (see Table~1 in Section~3).
Using the combination law
(\ref{eqT6}), one arrives at
$\Delta z_k\approx\Delta^{\!\rm syst}z_k\approx 1.5~$nm.

\subsection{Two Methods of Comparison between Experiment and Theory}

We begin with the first method where the computed theoretical Casimir forces of
Sections~4 and 5
are directly compared with the experimental estimations of the true force values giving
due account to their errors and uncertainties.

\begin{figure}[!t]
\centering
\includegraphics[width=.8\columnwidth]{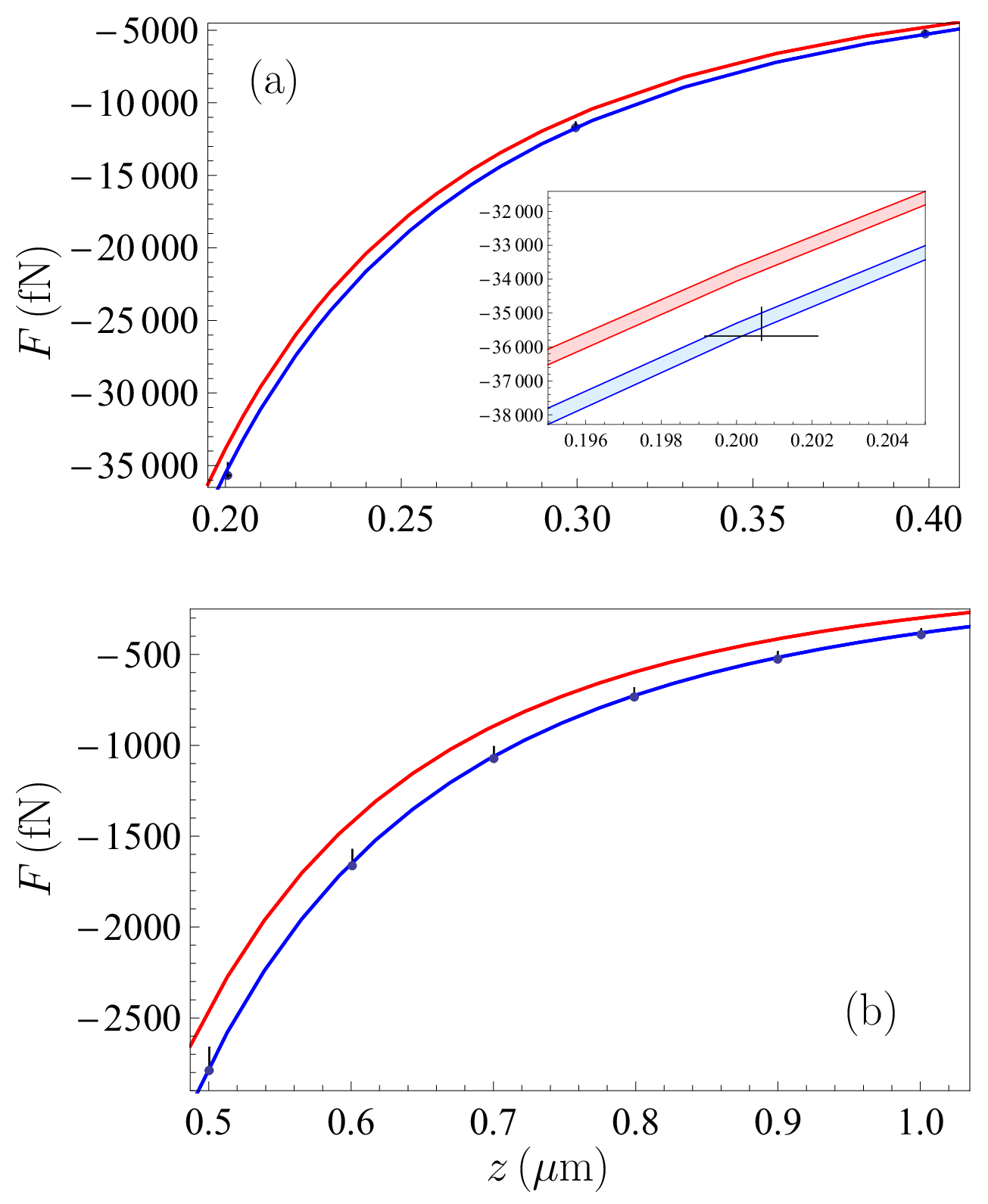}
\caption{The theoretical Casimir forces computed using  extrapolations of the optical
data by the Drude and the plasma
models are shown as  functions of separation by the red and blue
lines, respectively, over the region of separations (a) from 0.2 to $0.4~\mu$m and
(b) from 0.5 to $1.0~\mu$m. The experimental Casimir forces are indicated as
 crosses.
 (a) For better visualization the first cross at $z_1=200.6~$nm separation is shown
 on an enlarged scale.
 \label{figT4}}
\end{figure}

\begin{figure}[!t]
\centering
\includegraphics[width=.8\columnwidth]{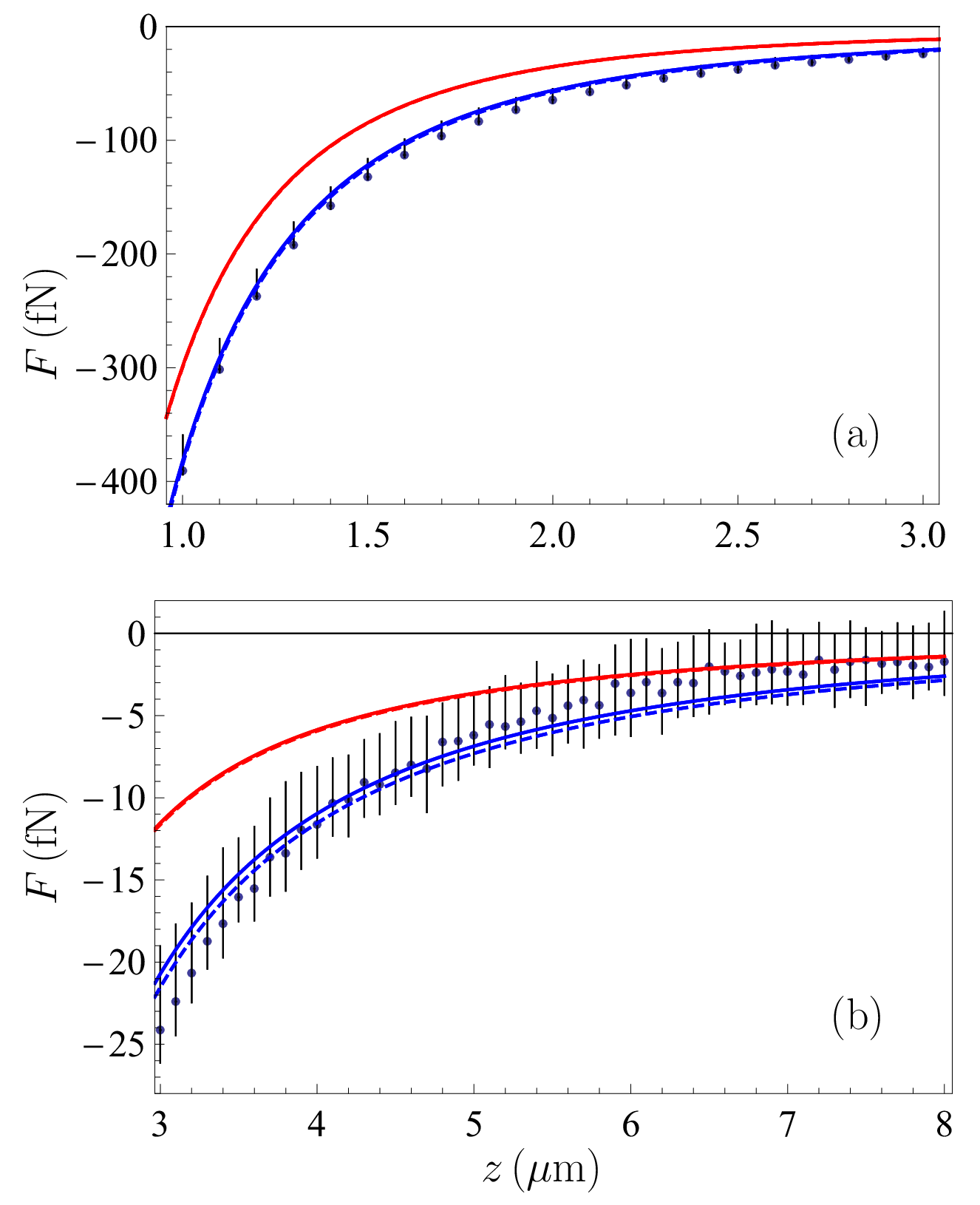}
\caption{The theoretical Casimir forces computed exactly using  extrapolations of the optical
data by the Drude and the plasma
models are shown as  functions of separation by the red and blue solid
lines, respectively, over the region of separations (a) from 1.0 to $3.0~\mu$m and
 (b) from 3.0 to $8.0~\mu$m.
Similar results obtained using the proximity force approximation are shown by the
red and blue dashed lines.
The experimental Casimir forces are indicated as
 crosses.
 \label{figT5}}
\end{figure}

In Figures~\ref{figT4}(a,b) and \ref{figT5}(a,b), the theoretical Casimir forces
computed in the
experimental configuration at $T=295.25~$K, as described in Sections~4 and 5 using the
Drude and plasma model extrapolations of the optical data, are shown  by the
solid red and blue
lines, respectively,  within the separation regions from 200~nm to $1~\mu$m
and from $1~\mu$m to $8~\mu$m, respectively.
In the same figures, the measured Casimir forces with their experimental errors and
uncertainties, including the role of surface patches,
are presented as crosses whose meaning is explained in Figure~\ref{figT3}.
An inset in Figure~\ref{figT4}(a) shows the first cross on an enlarged scale.

We recall that the experimental data shown in Figures~\ref{figT4} and \ref{figT5}
are obtained from the first harmonic of the signal. We have checked that they are
in good agreement with the force data obtained from the analysis of 21 harmonics
considered in Section~3.2 as the best fit of the Fourier coefficients.
As one example, the estimation of the true force value at $z_1=200.6~$nm shown
on the inset to Figures~\ref{figT4}(a) is given by
$\tilde{F}(z_1)=(-35683\pm 120)~$fN. This is in agreement with the value
(\ref{estfor}) found in Section~3.2. Note that the error indicated in (\ref{estfor})
is determined by inaccuracies of the fit and does not include the systematic error
of force measurements equal to 85.8~fN at this separation.

As is seen in Figures~\ref{figT4} and \ref{figT5}, an approach to calculation of the
Casimir force using the Drude model
is excluded by the data at the 95\% confidence level over the wide
region of separations from 200~nm to $4.8~\mu$m (in previous measurements
\cite{8,9,10,11,13,14,17,19,20} the Drude model approach was
experimentally excluded over the separation region from 162~nm to $1.1~\mu$m).
The plasma model approach is consistent with the data over the entire measurement
range. Note that the width
of theoretical lines is determined with account of the 0.5\% error in the force values due to
errors in the optical data of Au and errors resulting from the $0.2~\mu$m error
in the sphere radius.

In Figure~\ref{figT5}, we also show theoretical predictions of the standard Lifshitz
theory in the plane parallel geometry and the proximity force approximation
(see, e.g., \cite{ourBook,T4}) used
previously for the comparison between experiment and theory (red and blue dashed lines
obtained with extrapolations of the optical data by means of the Drude and plasma
models, respectively). It is seen that for the Drude extrapolation the exact results
are almost coincident with those obtained using the proximity force approximation.
For the plasma-model extrapolation, there are, however, some visible deviations
between the two sets of results which remain in the limits of experimental errors
[see Figure~\ref{figT5}(b)].

We continue with one more method for presentation of the measurement
data and their comparison with theory. Within this method \cite{ourBook,T4}, an
estimation of the true value of the Casimir force, $\tilde{F}^{\,\rm expt}(z_k)$,
measured at the separation $z_k$, is compared with the theoretical value $F^{\,\rm th}(z_k)$.
This is made by calculating and plotting in a figure as dots the force differences
\begin{equation}
F^{\,\rm th}(z_k)-\tilde{F}^{\,\rm expt}(z_k),
\label{eqT7}
\end{equation}
\noindent
where $F^{\,\rm th}(z_k)$ are computed at the experimental separations $z_k$ by using
either the Drude or the plasma model extrapolations of the optical data and
$\tilde{F}^{\,\rm expt}(z_k)$ are
defined in (\ref{eqT2}). The error of the quantity (\ref{eqT7}) at the desired
confidence level $\beta$ is given by
\begin{equation}
\Delta\left[F^{\,\rm th}(z_k)-\tilde{F}^{\,\rm expt}(z_k)\right]=
\Delta F^{\,\rm th}(z_k)+\Delta\tilde{F}^{\,\rm expt}(z_k).
\label{eqT8}
\end{equation}

The total experimental error $\Delta\tilde{F}^{\,\rm expt}(z_k)$ at different
separations was found in (\ref{eqT7a}) and shown by the pink dots in Figure~\ref{figT2}.
The total theoretical error  $\Delta F^{\,\rm th}(z_k)$ in (\ref{eqT8})
 is the combination of  errors in the calculated force values due to the error
in the sphere radius, errors in the optical data
of Au, already discussed above, and the error in separations $\Delta z_k=1.5~$nm.
The error in theoretical force values arising from the error in separations is
specific only for the second method of comparison between experiment and theory
because in this case the force values are calculated not over the separation
interval from 200~nm to $8~\mu$m, as in the first method, but only at the discrete
experimental separations determined with the error of 1.5~nm. These three
theoretical errors are combined by the law (\ref{eqT6}) with $k_{0.95}(3)=1.11$.
The obtained values of $\Delta F^{\,\rm th}(z_k)$ are
900, 32, 2.9, and 0.8~fN at 0.2, 0.5, 1.0, and $1.5~\mu$m, respectively.
At $z_k> 2~\mu$m we have
$\Delta F^{\,\rm th}(z_k)\ll\Delta\tilde{F}^{\,\rm expt}(z_k)$, so that the theoretical
error does not influence  the error of force difference (\ref{eqT8}).

\begin{figure}[!b]
\centering
\includegraphics[width=.8\columnwidth]{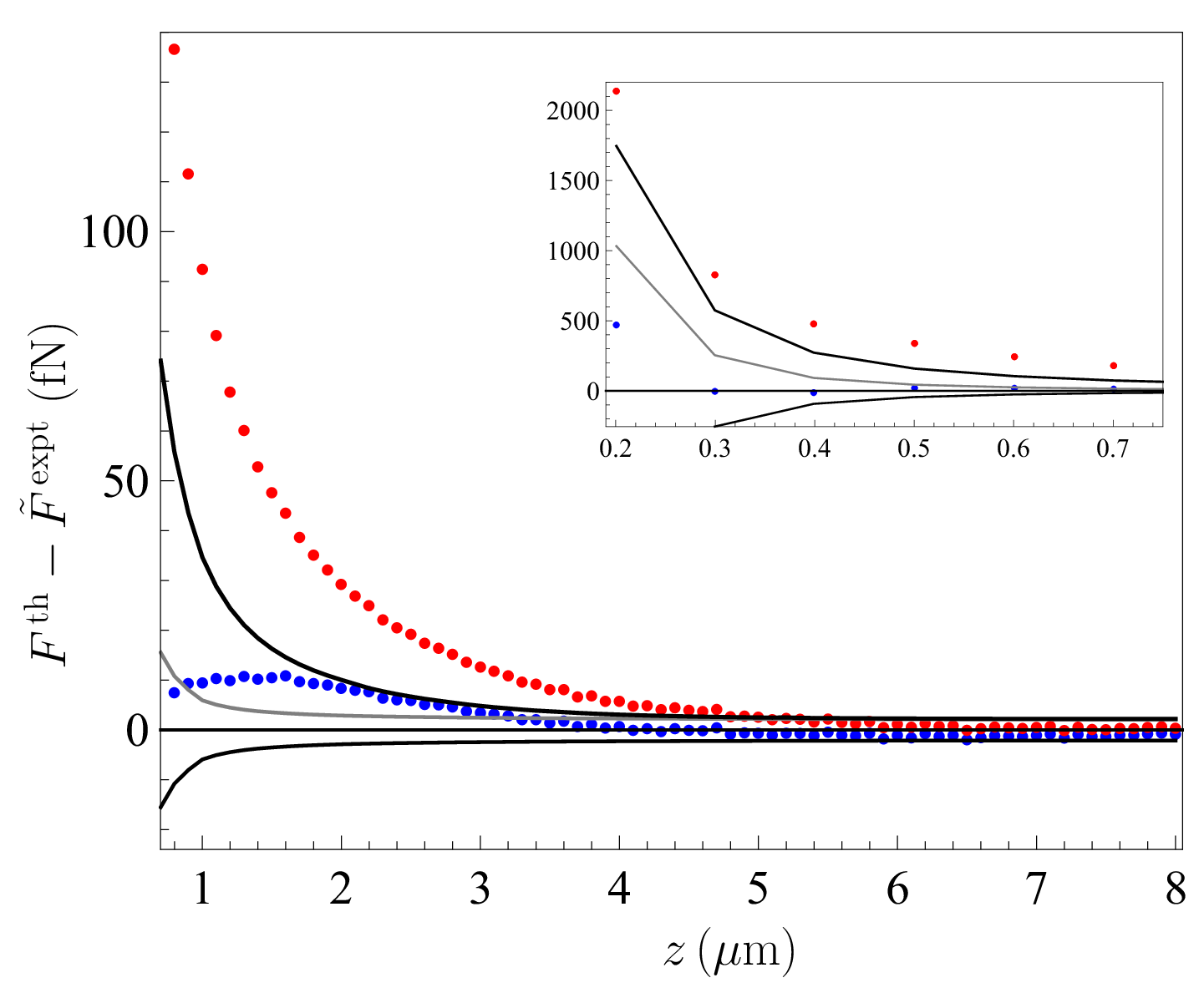}
\caption{The differences between theoretical, computed using the Drude and plasma model
extrapolations of the optical data,
and experimental Casimir forces are shown by the red and blue dots,
respectively, as the functions of
separation. The black solid lines indicate the borders of the confidence band
found at the 95\% confidence level. The gray line shows the upper border of the confidence
band as it would be in the absence of patches.
The region of separations below $0.8~\mu$m
is shown on an inset.
\label{figT6}}
\end{figure}
In Figure~\ref{figT6}, we plot the force differences (\ref{eqT7}) computed in the
separation region from 800~nm to $8~\mu$m using the Drude (red dots) and the plasma
(blue dots) model extrapolations of the optical data.
In the inset to  Figure~\ref{figT6}, the same is done in the separation
region from 200~nm to 700~nm. The two black lines indicate the borders of the 95\%
confidence band for the force differences. These borders consist of the ends of
segments
\begin{equation}
\left[-\Delta\Big(F^{\,\rm th}(z_k)-\tilde{F}^{\,\rm expt}(z_k)\Big),\,
\Delta\Big(F^{\,\rm th}(z_k)-\tilde{F}^{\,\rm expt}(z_k)\Big)+
|F_{\rm patch}(z_k)|\right]
\label{eqT9}
\end{equation}
\noindent
computed using (\ref{eqT8}) and linked together by the straight lines.
The right ends of the confidence intervals (\ref{eqT9}) include a contribution
originating from the electrostatic patches (see Figure~\ref{figPatch} in
Section~3). For a clearer understanding of their role, we also plot by the gray
line the upper border of the confidence interval as it would be in the absence
of patches.
As is seen in Figure~\ref{figT6}, the theoretical approach using the Drude model
is excluded by the
measurement data within the separation region from 200~nm to $4.8~\mu$m,
where all the red dots are outside the confidence band. This conclusion is in line
with that obtained using the first method of comparison between
experiment and theory.

In a similar way,
as it is seen in Figure~\ref{figT6}, the plasma model approach is consistent with the
measurement data over the entire measurement range.
This is again in agreement with the conclusion made above.

\section{Discussion}

In the foregoing, we have presented the results of an
experiment measuring the differential Casimir force
between an Au-coated sphere and top and bottom of deep
Au-coated trenches concentrically located on a rotating
disc performed by means of micromechanical torsional
oscillator. Due to a sufficiently large deepness of these
trenches, the measured force signal follows the Heaviside
function, i.e., the trench bottoms do not contribute to
the force. Thus, this experiment measures directly the
Casimir force between a sphere and a plane plate
simultaneously preserving all the advantages of differential
force measurements including the highest level of sensitivity.
This allowed obtaining the meaningful experimental results
at separations up to a few micrometers and distinguish
between different theoretical predictions for the Casimir
force.

To reach this goal, it was necessary to analyze the
distribution laws of the measurement data in order to
find reliable estimations of the true force values and
confidence intervals at each separation and collect
together all sources of the systematic errors. As a result,
the total experimental error was found at the 95\%
confidence level as a function of separation. In order
to reveal all factors which could make an impact on the
comparison between experiment and theory, the roles of
surface roughness and edge effects have been carefully
investigated and found to be negligibly small. The
interacting surface was characterized by Kelvin probe
microscopy which gave the possibility to estimate the
typical size of surface patches, the r.m.s. voltage
and related uncertainties in the measured force. These
uncertainties were taken into account in the error
analysis along with the random and systematic errors.

The theoretical Casimir force between the Au-coated
surfaces of a sphere and a plate used in the experiment
were calculated on the basis of first principles of
quantum electrodynamics at nonzero temperature by means
of the scattering theory and the gradient expansion
without resort to any simplified approximations of
additive character. The results obtained within these
two approaches were found to be in excellent agreement.
The computations were done using the optical data of Au
extrapolated to zero frequency by means of the Drude
and plasma models.

Direct comparison between experiment and theory with
no fitting parameters demonstrates that an extrapolation
by means of the Drude model is excluded by the measurement
data over the separation region from 0.2 to 4.8 $\mu$m,
whereas the theoretical predictions using the plasma model
are experimentally consistent over the entire measurement
range. This significantly widens the range of separations
where the theoretical approach using the Drude model was
excluded so far and again raises a question on the
physical reasons of this result.

In fact the relaxation properties of conduction electrons
taken into account by the Drude model and omitted by the
plasma one do exist and are observed in numerous physical
phenomena other than the Casimir effect. Because of this,
future theory of the Casimir force must take them into
account in one way or another. An attempt in this
direction is already undertaken \cite{last}. The authors
hope that the solution to this problem will be found in a
not too remote future.

\section{Conclusions}

To conclude, the performed experiment on measuring the
Casimir force between an Au-coated surfaces of a sphere
and a plate by means of a micromechanical torsional
oscillator reached an unprecedented precision in the
wide separation region from 0.2 to 8~$\mu$m. A
comparison of the obtained measurement data with the
exact theory based on both the scattering approach and
the gradient expansion with no fitting parameters allowed
clear discrimination between the theoretical predictions
using the Drude and plasma model extrapolations of the
optical data up to an unusually large separation distance
of 4.8~$\mu$m. As a result, the Drude extrapolation was
excluded by the measurement data, whereas the
extrapolation using the plasma model turned out to be
consistent with the data. At the moment the generally
recognized physical explanation for these facts is
lacking.

\funding{P.A.M.N. was supported by the Brazilian agencies National
Council for Scientific and Technological Development (CNPq), Coordination
for the Improvement of Higher Education Personnel (CAPES), the National
Institute of Science and Technology Complex Fluids (INCT-FCx), and the
Research Foundations of the States of Rio de Janeiro (FAPERJ) and
S\~{a}o Paulo (FAPESP).
The work of G.L.K.~and V.M.M.~was supported by the Peter the Great
Saint Petersburg Polytechnic
University in the framework of the Russian state assignment for basic research
(project N FSEG-2020-0024).
V.M.M.~was also partially funded by the Russian Foundation for Basic Research grant number
19-02-00453 A.
R.S.D.~was partially funded by National
Science Foundation through grant PHY1707985}

\acknowledgments{
P.A.M.N.~thanks IUPUI for hospitality during his stay in Indianapolis.
V.M.M.~is grateful for partial support by the Russian Government Program of Competitive
Growth of Kazan Federal University. R.S.D.~acknowledges financial
and technical support from the IUPUI Nanoscale Imaging
Center, the IUPUI Integrated Nanosystems Development
Institute, and the Indiana University Center for Space
Symmetries. }

\reftitle{References}

\end{document}